\documentclass[]{pasj02} 
\usepackage[switch,mathlines]{lineno} 
\usepackage{graphicx}
\usepackage{enumerate}
\usepackage{multirow}
\usepackage{diagbox}

\jyear{2026}
\Received{2026/03/23}
\Accepted{2026/09/28}

\graphicspath{{./}{figures/}} 

\begin{document} 

\title{ Survey of small amplitude waves in solar prominences }

\author{
Kiyoshi \textsc{Ichimoto},\altaffilmark{1,2}\altemailmark\orcid{0000-0001-8689-3564} \email{ichimoto@kwasan.kyoto-u.ac.jp} 
Ayumi  \textsc{Asai},\altaffilmark{2}\orcid{0000-0002-5279-686X} 
Takako T. \textsc{Ishii}, \altaffilmark{2}\orcid{0009-0004-4110-388X} 
and 
Satoru \textsc{UeNo}\altaffilmark{2}\orcid{0000-0002-2165-7998} 
}
\altaffiltext{1}{College of Science and Engineering, Ritsumeikan University, Kusatsu, Shiga, Japan}
\altaffiltext{2}{Astronomical Observatory, Kyoto University, Kyoto, Japan}




\maketitle

\begin{abstract}
Presence of small amplitude oscillation or waves in solar prominences was investigated
by using archived data from a full disk H$\alpha$ imaging spectroscopic observation.
A total of 178 prominences on the limb were studied and it was found that there are plenty of evidence 
of oscillations or waves with preferential periods in 3 -- 5 min (4 min) and 10 -- 20 min (15 min).
We found very clear signature of persistently propagating 4 min wave in 21 prominences
and at least a hint of 4 min oscillation in 90 prominences in their Doppler signal.
The 4 min waves are often found in prominences with a large horizontal size, i.e.,
long filaments seen from a side, and 
they propagate in nearly horizontal direction with a phase speed of 140 -- 400 km/s. 
On the other hand, very clear evidence of the 15 min wave was found
in 14 prominences and at least a hint of oscillation in 62 prominences in the Doppler signal.
The 15 min waves are preferentially propagate in vertical direction with a phase speed of 20 -- 50 km/s.
There is a positive correlation between the presences of 4 min and 15 min waves.
The 4 min waves are interpreted as Alfv\'{e}nic wave propagating along the coronal magnetic field supporting the prominences.
Two scenarios are discussed for the 15 min wave, i.e., 
Alfv\'{e}nic wave propagating vertically in prominence bodies 
and apparent cross-field propagation of ensemble of Alfv\'{e}nic waves in magnetic structures.
Prominences that harbor the short period waves or oscillations distribute over the entire solar latitude,
suggesting that these waves are excited by ubiquitous disturbance in solar atmosphere.
While it is also true that a number of prominences do not show signatures of oscillations 
nor propagating waves, and the reason of this diversity remains open question.
Our study revealed that persistently propagating short period waves in prominences 
is not particularly rare but rather common phenomena,
and suggests that observing the 4 min waves in prominences will serve as
a useful tool for tracing the coronal Alfv\'{e}nic waves and the magnetic field supporting the prominences.
The result of this study implies that the magnetic fields are preferentially horizontal in long quiescent prominences.
\end{abstract}


\section{Introduction}\label{sec:1}

Solar prominences are cool and dense plasma supported  by the magnetic field
in surrounding hot and tenuous corona. 
As inferred from their organized shapes, 
prominences are basically low $\beta$ plasma whose structure and dynamics are governed by the
magnetic field.
In fact, they exhibit rich phenomena driven by the magnetic field and provide us an excellent laboratory
for studying the astronomical magneto-hydrodynamics (MHD).
Diagnosing the magnetic field of prominences is, therefore, an important observational task today,
and high accuracy spectro-polarimetric observations have been conducted for this purpose
by a number of researchers.
However, determining the magnetic field of prominences is still a challenge 
due to the weak polarization signals in spectral lines and 
ambiguities in determining the magnetic field vector using Zeeman and Hanle effects.
Especially, whether the magnetic field in a prominence is predominantly vertical or horizontal 
with respect to the solar surface is still contradictory or often uncertain
(e.g. \cite{Bommier_1994, Schmieder_2013, Yamasaki_2023, Merenda_2006, Gonzalez_2015, Hashimoto_2026}).

Oscillations or waves in solar prominences has been paid much attention 
not only for its interest as the fundamental MHD process but also 
in connection to the energy transport in prominences and corona \citep{Okamoto_2015, Hashimoto_2023},
and as a tool for seismological diagnosis for physical conditions of prominences including their magnetic fields
\citep{Oliver_2002, Ballester_2015}. 
Prominence oscillations are classified into two category by their amplitude, i.e., 
large amplitude oscillation with a velocity amplitude larger than 10 -- 20 km/s, 
and small amplitude oscillation with a velocity amplitude smaller than $\sim$5 km/s 
\citep{Oliver_1999, Ballester_2015, Arregui_2018}.	
The former is impulsively excited by an energetic disturbance induced by a strong flare 
or jet \citep{Joshi_2023}
occurred in a distance from the prominence, 
while the excitation mechanism of the later is poorly known , i.e.,
they are observed frequently without noticeable external disturbances.
In this paper we focus on the short period small amplitude oscillation that is likely more ubiquitous
and then preferable for a statistical study for diagnosing the prominences.
Comprehensive reviews of past studies on the small amplitude oscillation of prominences are found in 
\citet{Tsubaki_1988}, \citet{Oliver_2009} and \citet{Arregui_2018}.

\begin{table*}
\caption{Past studies on short period (P$<$20min) small amplitude oscillations}\label{tab:papers} 
\begin{center}
\begin{tabular}{llcccccccl}
\hline
Author, year		& line	& signal$^1$ & period  & amplitude & target$^2$ & location & duration & phase speed\\
 	 			&  [nm]	&  		    & [min]   & [km/s]     &          &            &    [min]   &  $v_p$ [km/s] \\
\hline
\cite{Wiehr_1984} 	& H$\alpha$	& V 	& 3, 5 		& 1 -- 2	& P$\times$3 & - & - & - \\
\cite{Tsubaki_1986} 	& CaII K 		& V 	& 3.5 -- 4 	& $\le$2 		& QP & edge & - & - \\
\cite{Balthasar_1986} & H$\alpha$	& V 	& 3.6 -- 6.7 & $\sim$2 	& P & - & - & -\\
\cite{Tsubaki_1986} 	& CaII K 		& V 	& 10 		& 1.2 -- 3 	& QP & local ~5Mm & - & - \\
\cite{Suematsu_1990} 	& CaII K, H$\beta$ & V & 4, 14 & $\sim$0.5 	& QP & edge & - & - \\
\cite{Yi_1991a}		& HeI 1083	& V,I & 5, 9, 16 	& $\sim$0.5 	& QF & threads & - & $\sim$30\\
\cite{Yi_1991b}		& HeI 1083	& V,I & 9 -- 16 	& $\sim$0.5 	& QF  & threads & - & $\sim$30\\
\cite{Thompson_1991} & H$\alpha$ & V 	& 2.5 -- 4.3 	& $\sim$1       & F & - & - & $150\sim200$ \\
\cite{Balthasar_1993} & CaII 854, HeI D3 & V & 0.5, 12, 20& $\sim$2 & P & a few asec & - & - \\
\cite{Molowny-Horas_1997} & H$\beta$  & V & 7.5        &       -      &  QP & - & $\sim$12 & >4.4\\
\cite{Suetterlin_1997} & CaII 854 & V & 20       &       -      &  P$\times$8 & a few asec & a few cycles & - \\
\cite{Okamoto_2007} & CaII H 		& T 	& 2.8 -- 4.3 	& $\sim$10 & AP & threads & - & >1050\\
\cite{Lin_2007} & H$\alpha$	& V 	& 3 -- 9 	& 1 -- 2 & QF & threads & - & $\sim$12 \\
\cite{Berger_2008} & CaII H, H$\alpha$	& T 	& 20 -- 40 	& - & QP & - & - & $\sim$10 \\
\cite{Ning_2009} 	& H$\alpha$ 	& T 	& 3.5 -- 8.8 	& $\sim$3.5 & QP & local & 1-2 cycles & - \\
\cite{Schmieder_2013} 	& CaII H & T & 1 -- 2 	& - & QP & - & - & 10$\sim$15 \\
\cite{Hillier_2013}	& CaII H, H$\beta$ & I, V & $\sim$	5 	& 1 -- 4 & QP & ubiquitous & episodic & - \\
\cite{Orozco_2014} 	& HeI 1083nm 	& V 	& $\sim$3.7 	& $\sim$0.3 & QP & local & 6.7 & - \\
\cite{Okamoto_2015} 	& CaII H/MgII h 	& T, V & $\sim$5 	& $\sim$10 & AP & threads & damp & -\\
\cite{Ofman_2015} 	& CaII H, H$\beta$ & I, V & 5 -- 11 	&  - & QP & - & - & 5.9--8.5 \\
\cite{Khomenko_2016} &  CaII854, HeI1083 & V & $\sim$3 & - & QP & whole & >30 & - \\
\cite{Li_2022}   & H$\alpha$     &T, V &	10 -- 16 & $\sim$4 &	QP	& thread & persistent & $\sim$160\\
\cite{Song_2024} & H$\alpha$ & I$^3$ & 4 -- 7, 9 -- 12 &  -  & QP & threads & persistent & - \\
\cite{Ichimoto_2024} & HeI 1083, H$\alpha$ & V & $\sim$4, 15 & $\sim$2 & QP & whole & persistent & 240, 27 \\
\hline
\end{tabular}
\end{center}
\begin{tabnote}
\footnotemark{} Signal of oscillation detected in LOS velocity (V), intensity (I). transversal motion (T).
\footnotemark{} QP: quiescent prominence, AP: active region prominence, P: either quiescent or active, QF: quiescent. \footnotemark{} Intensity at single wavelength point in H$\alpha$ that may be affected by Doppler shift. 
\citet{Song_2024} also found transversal oscillations with a period of 10 - 50 min..
\end{tabnote}
\end{table*}

Table 1 summarizes past studies on the short period small amplitude oscillations.
Here we refer the short period as a period shorter than $\sim$20min,
and the table does not include the past reports of waves with a longer period.
Since its early observations in 1980th \citep{Wiehr_1984, Tsubaki_1986, Balthasar_1986}, 
the small amplitude oscillation has been reported by a number of observers.
They are usually observed in a limited area in
prominences preferentially around the edge of them, 
and they are episodic in time with a damping in a few cycles \citep{Oliver_2009, Arregui_2018}.
Exceptionally \citet{Khomenko_2016} reported persistent oscillations in whole prominence
with a period of around 3min.
while  \citet{Li_2022} reported a persistent propagating wave in a quiescent prominence with a period and 
propagating speed of 10 -- 16min and 167km/s, respectively.
As seen in Table 1, some authors reported the phase speed of the waves, i.e., 
their propagating nature, 
while others not, possibly interpreted oscillations as standing waves in prominences
representing their eigenfrequencies.
Periods of oscillation are different among the observations,
but there seems to be two preferential ranges, i.e.,  3 -- 6 min and 10 -- 20min (see Table 1).

Recently \citet{Ichimoto_2024} (hereafter Paper I) reported a presence of persistent waves propagating across 
a quiescent prominence from a spectro-polarimetric observation in HeI 10830\AA.
Dominant frequencies are about 4 min and 15 min and they propagates preferentially in horizontal and vertical
directions with a phase speed of approximately 140 -- 200 km/s and $\sim$24 km/s, respectively.
These waves were found only in the line-of-sight (LOS) velocity but not in the intensity of the spectral line.
With the magnetic field strength obtained from the spectro-polarimetric data, 
they concluded that the 4 min oscillation is an Alfv\'{e}nic wave propagating 
in the corona harboring the prominence threads,
and the 15 min oscillation is Alfv\'{e}nic wave propagating upward in the prominence body.
This finding suggests that the waves in prominences with a period around  4 min will provide a mean
to diagnose the magnetic field supporting the prominences in the corona.

.
Detection of such persistent small amplitude waves propagating across the
entire body of prominences is quite rare in past observations as seen in Table 1,
and it is of a great interest if such oscillations or waves are really uncommon phenomena or not.
Remarkably the waves found in Paper I were clearly seen also in the data 
from a fulldisk H$\alpha$ imaging observation in the same prominence.
This paper aims to investigate the presence of small amplitude waves for a large number of prominences
using a synoptic H$\alpha$ imaging observation and to test the universality of such waves.
Our motivation is also in diagnosing the magnetic field supporting the prominence 
with the idea that, 
if a propagating Alfv\'{e}nic wave is detected,  the direction of its propagation
directly infers the direction of the magnetic field penetrating the prominences.
In the following, we describe the observational dataset and method of data reduction in section 2,  
results in section 3, and discussion and summary in section 4.

\begin{figure*}[t]
\centerline{
	\includegraphics[width=0.8\linewidth]{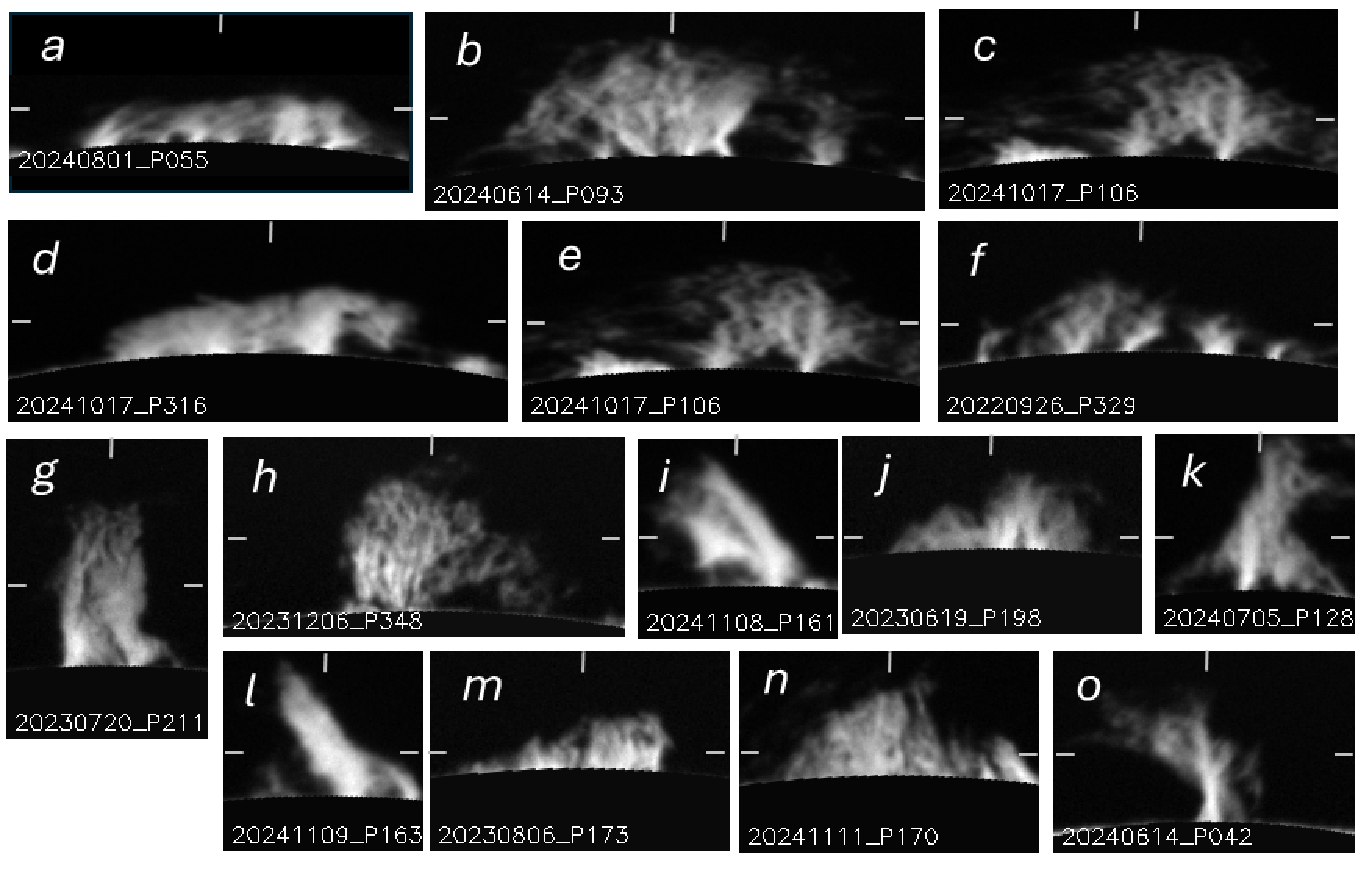}
}
\caption{Examples of prominences in our dataset. Images are in H$\alpha$ center.
Positions of the central slits in horizontal and vertical directions used to 
produce the space-time plots are indicated by tick marks in both sides and at the top
of prominences, respectively.
{Alt text: 15 images of prominences taken in H$\alpha$ are displayed.
The bottom of each panel is the solar limb with the disk below which is masked.
Each image is labeled by  $a$ -- $o$.}
}
\label{fig_images}
\end{figure*}

\section{Dataset and analysis}\label{sec:2}

\subsection{Data selection}
Dataset used in this study was obtained by the Solar Dynamics Doppler Imager 
(SDDI, \citet{ichimoto_2017}) on the Solar Magnetic Activity Research Telescope
 (SMART, \citet{Ueno_2004}) at Hida Observatory of Kyoto University in Japan.
SDDI has been regularly taking full disk solar images through a tunable Lyot filter
at 73 wavelength positions around H$\alpha$ with a $\lambda$-step of 0.25\AA\ since 2016.
The spatial sampling and time cadence are 1.23''/pixel and 12 sec, respectively.
Prominences investigated in this study are selected with the following criteria
from the archived data of the SDDI; 
the data has continuous time coverage longer than 2.5 hours 
and the prominence has a size larger than 20 arcsec in its apparent length and height
so that we are able to detect a propagating wave if it is present.
Activated prominences showing a significant change in their shape during the observation
and data suffering a significant noise due to a poor sky condition are excluded.
As a result, we selected 178 prominences on the limb from the archived data 
recorded in the period from 2022 May to 2024 November.
Our sample  includes 169 quiescent prominences, in which 49 are polar crown prominences, 
3 active region prominences and 6 intermidiate prominences.
The reason of the small number of active region prominences is likely due to their short height
on the limb.
To give a rough idea on our prominence dataset and display the variety of their shape, 
some examples of H$\alpha$ images are presented in figure \ref{fig_images}.
This sample contains prominences that exhibit clear oscillations or waves and prominences 
that do not, which will be discussed in the later section. 

\begin{figure*}[h]
\centerline{
	\includegraphics[width=1.0\linewidth]{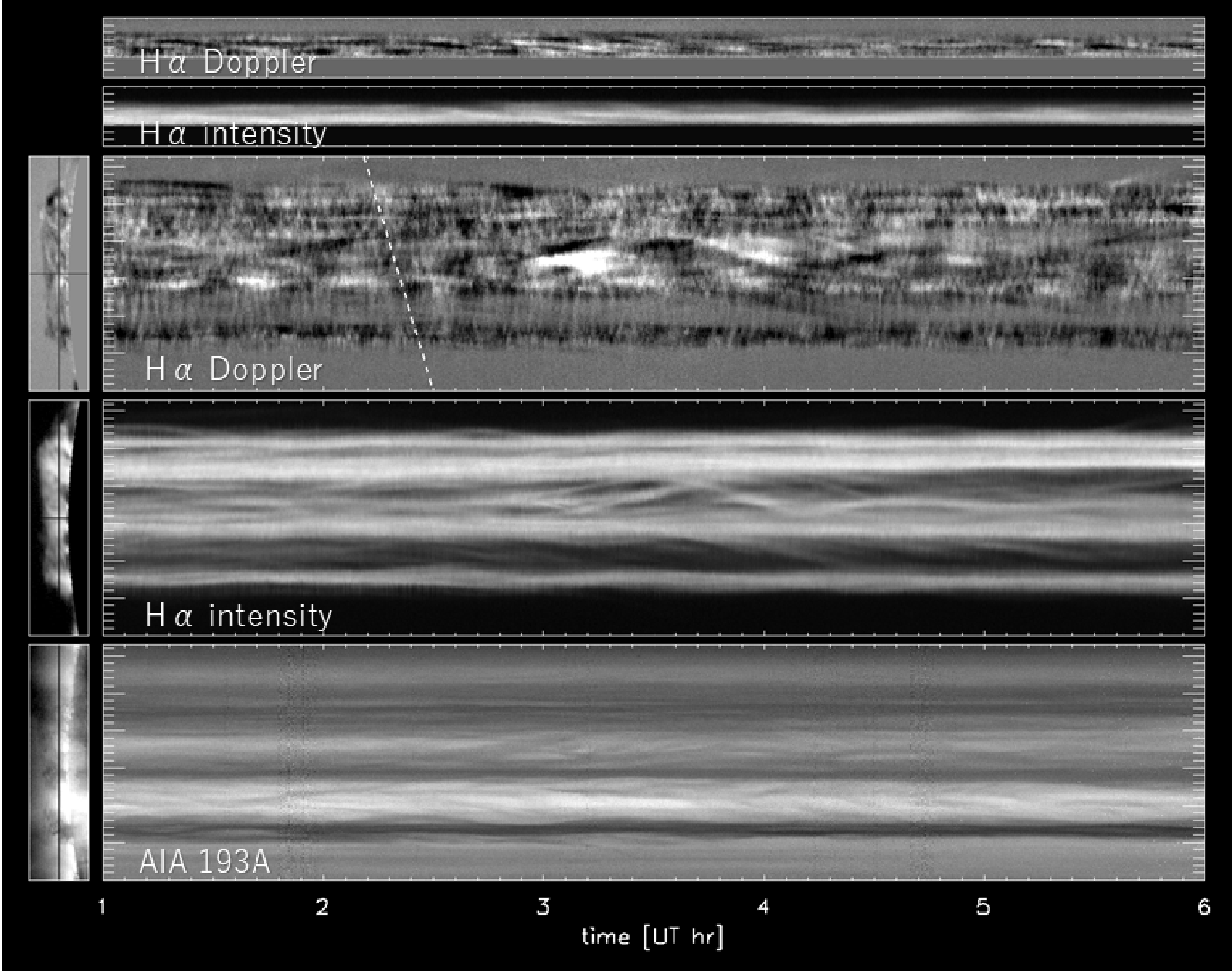}
}
\caption{Space-time plot for the prominence on 2024 August 1  (panel $a$ of Fig. \ref{fig_images}). 
The horizontal axis is time and vertical axes are position along the central slits.
From the top to the bottom, H$\alpha$ Doppler signal and H$\alpha$ intensity along the vertical slit, 
H$\alpha$ Doppler signal, H$\alpha$ intensity and AIA 193A intensity along the horizontal slit, respectively.
The lengths of the vertical and horizontal slits are 80 and 315 arcsec, respectively, 
and each tick of the vertical axes corresponds to 10 arcsec.
On the left of lower 3 panels shown are corresponding images of the prominence rotated by 90$\degree$,
where the positions of horizontal and vertical slits are displayed.
Note that the prominence is seen as dark features in the AIA image.
The oblique dashed line in the 3rd panel indicates the slope of the phase velocity of 200 km/s.
{Alt text: Five gray scaled panels are arranged vertically.
The x-axis is common for five panels representing the time in hour in UT spanning from 1 to 6.
The y-axes are distances along the slits crossing the prominence.
The gray scale of each panel presents, from the top to the bottom, 
H$\alpha$ Doppler signal and H$\alpha$ intensity along the vertical slit, 
H$\alpha$ Doppler signal, H$\alpha$ intensity and AIA 193A intensity along the horizontal slit, respectively.
On the left of lower 3 panels shown are images of the prominence.
Dense stripes or oscillatory patterns are clearly seen in the third panel.}
}
\label{fig_timeslice}
\end{figure*}

\subsection{Data reduction}

After dark and flat field corrections of the fulldisk data, 
we performed the following processes for each prominence.

\begin{figure}[h]
\centerline{
	\includegraphics[width=1.1\linewidth]{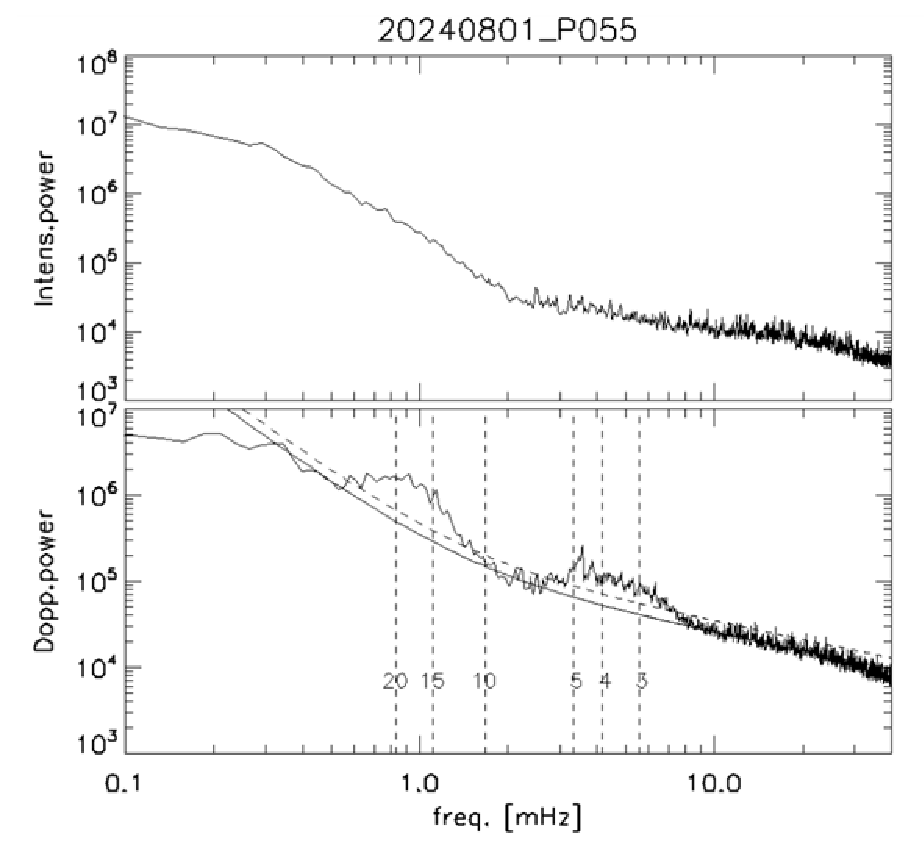}
}
\caption{Averaged power spectral density of H$\alpha$ intensity (top) and Doppler signal (bottom)
for the same prominence of figure \ref{fig_timeslice}.
The solid curve is the base of the power spectrum represented by a combination of
two red noises ($\propto f^{-\gamma}$) with the index $\gamma$ of 2.5 and 0.7.
Shown by the dashed curve is the 95\% confidence level calculated by assuming 
a $\chi^2$ distribution of the power spectra with 100 degrees of freedom.
{Alt text: Two graphs are shown one above the other.
The common x-axis is frequency in Hz in logarithmic scale ranging from 0.001 to 0.04.
The y-axes of upper and lower panels are powers spectrum density of intensity and Doppler velocity,
respectively.
Particular frequencies corresponding to periods of 3, 4, 5, 10, 15, 20 min are shown 
by vertical dashed lines in the lower panel.
In the lower panel, two enhancements of the power are recognized around 4 min and 20 min.}
}
\label{fig_power}
\end{figure}

\begin{figure*}[h]
\centerline{
	\includegraphics[width=0.8\linewidth]{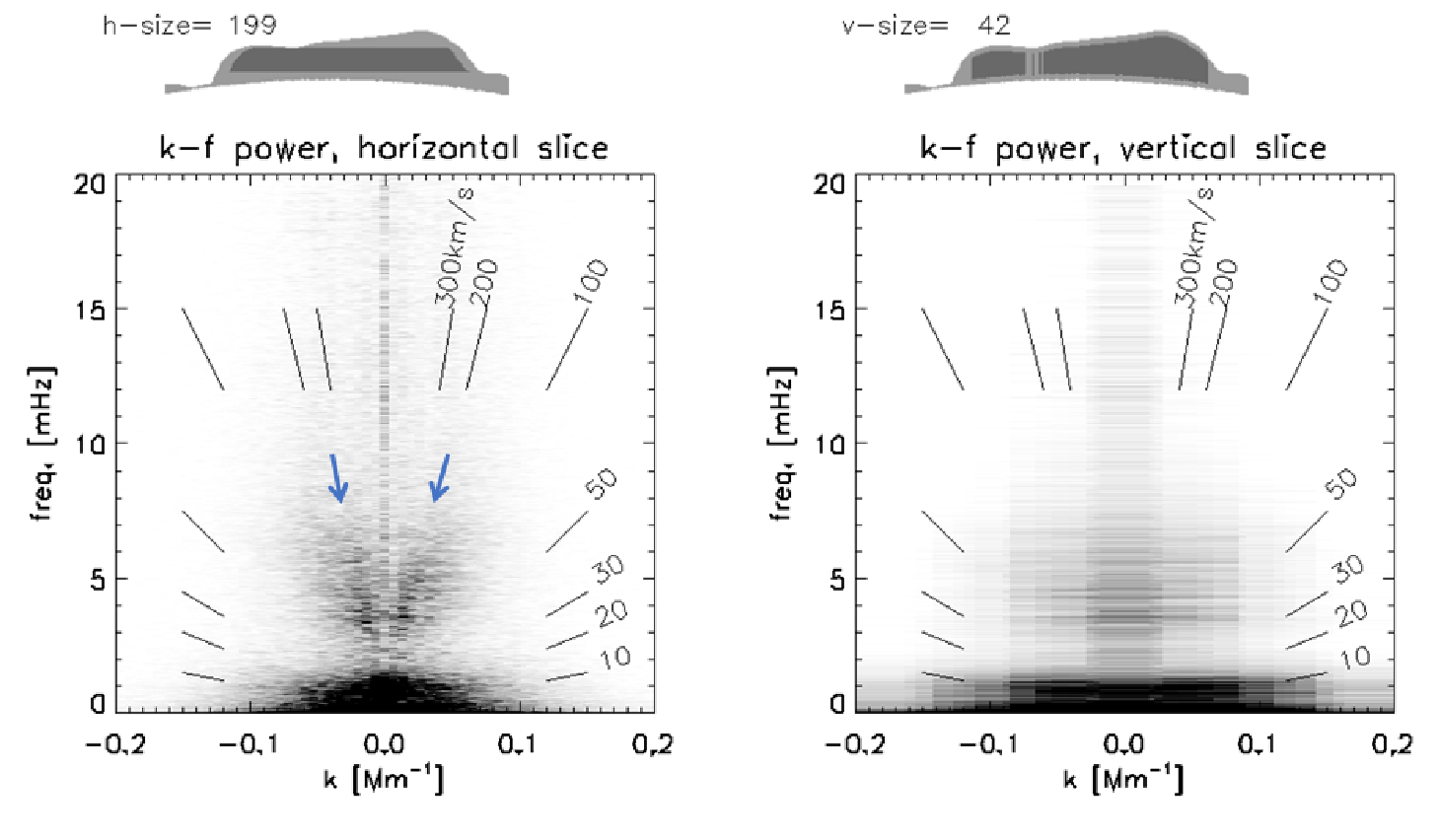}
}
\caption{Power spectrum density of the LOS velocity on $(k-f)$ planes 
for the same prominence of figure \ref{fig_timeslice},
where $k$ is the wavenumber along horizontal (bottom left panel) 
and vertical (bottom right panel) slits, respectively, and $f$ is the frequency in Hz.
Two arrows in the left panel indicate the oblique ridges of enhanced power.
At the top, dark parts in prominence show the area filled with horizontal (left) and vertical (right)
slits by which horizontal and vertical sizes of the prominence (shown at the top in pixels)
are measured, respectively.
{Alt text: Gray scale 2D maps in the left and right.
The x-axis of each panel is wave number $k$ in per Mm ranging from $-0.2$ to $0.2$,
and the y-axis is frequency $f$ in Hz ranging from 0.0 to 0.02.
At the top of each panel, the shape of prominence is shown by gray color.}
}
\label{fig_kfpower}
\end{figure*}

\begin{enumerate}[(1)]
\item  Extract partial images that contain the target prominence
from the fulldisk images in wavelength range from $-1.25$\AA\ to $+1.25$\AA\ . 
\item Identify the solar limb and the sky area (= outside the limb) in partial frame 
using a near continuum image (average of  $\pm$1.25\AA).
To avoid the chromosphere on the limb seen in H$\alpha$, we mask region below the height of 10 arcsec 
from the continuum limb.
\item Make sky spectrum by averaging darkest 100 points in H$\alpha$ central intensity
in the sky area.
\item Subtract the sky spectrum from the data for all pixels outside the limb.
\item Make Dopplergrams (LOS velocity maps) by subtracting red wing intensity 
(sum of $+0.25$ and $+0.5$\AA) from blue wing intensity (sum of $-0.25$ and $-0.5$\AA),
and H$\alpha$ intensity maps  (sum of $-0.25, 0,  +0.25$\AA) .
Here the bias in LOS velocity due to the offset of the central wavelength of the tunable filter 
is removed by removing the correlation between the LOS velocity and intensity.
\item Rotate the partial images by its polar angle to make the local limb horizontal and
at the bottom as seen in figure \ref{fig_images}.
\item Define the \it{central slits} \rm{} that cross the approximate center of the prominence 
and have horizontal (parallel to the nearby limb) and vertical directions, the position of which 
are shown by the white ticks in figure \ref{fig_images}.
\item Make space-time diagrams of the LOS velocity signal and H$\alpha$ intensity
for horizontal and vertical central slits.
An example, which is made for the prominence on 2024 August 1 (prominence \it{a} \rm{} in figure \ref{fig_images} ),
is shown in figure \ref{fig_timeslice}.
\item Define the area of the prominence by the pixels whose time-averaged H$\alpha$ intensity
is greater than 0.05 times of its maximum in the prominence.
\item Make power spectrum densities of the time variations 
of the LOS velocity and the intensity for each pixels in the prominence and average them.
Before the Fourier transform, a linear trend and erroneous data (spikes) are removed from each time series.
An example of the spatially averaged power spectrum density is shown in figure \ref{fig_power}.
\item Apply 2-dimensional Fourier transforms on the space-time diagrams 
of the LOS velocity along multiple horizontal and vertical slits inside the prominence, and 
make averaged $k-f$ diagrams of the power spectral densities for horizontal and vertical slits.
Here $k$ is the wave number along the slits in Mm$^{-1}$ and $f$ is the frequency in Hz.
The results are shown in figure \ref{fig_kfpower} for the same prominence as figure \ref{fig_timeslice}. 
Dark areas in the top left and right panels of figure \ref{fig_kfpower} depict the regions filled with the
horizontal and vertical slits, respectively.
\item Determine horizontal and vertical sizes of the prominence as the maximum length 
of the horizontal and vertical slits inside the prominence. 
On the top left and top right panels of figure \ref{fig_kfpower}, horizontal and vertical sizes of the
prominence are respectively shown in pixel.
\end{enumerate}

Note that a more relevant LOS velocity will be obtained by dividing the 
difference of intensities in red and blue wings by the sum of them or by 
fitting the spectral intensity profiles with a Gaussian.
However these methods are noise-sensitive and often introduce erroneous results 
especially in faint parts of prominences.
To conduct our analysis for a large amount of data in a robust way,
we used the simple difference of red and blue wing intensities 
as the proxy of the LOS velocity.
Therefore our power spectrum density of LOS velocity calculated from this proxy are
regarded as intensity-weighted average of the power of oscillations.
With a noise level of $\pm$(0.25+0.5)\AA\ intensities estimated in sky area, 
the sensitivity for the LOS velocity is evaluated as $\sim$0.5 km/s 
in prominences with a moderate H$\alpha$ intensity.

In addition to the size of prominences, 
we also measured the angle between the long axis of the prominences and
the line-of-site at the observation (LOS angle)
by identifying the corresponding dark filament on the solar disk 
seen in H$\alpha$ fulldisk images taken on proceeding or preceding days of observation 
of the prominence on the east or west limb, respectively.
The procedure for obtaining the LOS angle is described in Appendix A1.
The number of prominences for which we could obtain the LOS angle
is limited because it is sometimes impossible to identify the corresponding
dark filament on the disk due to their time evolution. 
Moreover, in case of high latitude prominences, 
it is difficulty to measure the angle due to their invariant closeness to the limb.
The number of prominences for which we could measure the LOS angle is 116 out of 178.
For the prominences that could be identified as a dark filament on the disk, 
we also obtained the longitudinal distance of the prominence from the limb at the time of observation.
This information will be used later to evaluate the real height of the prominences.

\subsection{Overview of the reduction products}
Here we revisit figures \ref{fig_timeslice} $\sim$ \ref{fig_kfpower}.
These plots are obtained for a prominence on 2024 August 1 at polar angle 55$\degree$
from the geographical north 
(panel \it{a} \rm{} of figure \ref{fig_images}), which is one of the prominences 
that showed conspicuous signatures of propagating waves and oscillations.
Remarkable features in the space-time diagram are oblique dens stripes seen in the LOS velocity 
for the horizontal slit (third panel from the top of figure \ref{fig_timeslice}).
These features indicate the presence of waves propagating in both leftward and rightward directions 
along the horizontal slit.
Approximate period and phase speed are 4 min and 200 km/s, respectively.
In addition to the 4 min periodicity, we also recognize oscillatory pattern with 3--4 cycles per hour
in the space-time diagram of the LOS velocity.
Contrary, such patterns are not recognized in space-time plot of the H$\alpha$ intensity
 (second and fourth panels from the top of figure \ref{fig_timeslice}).
Power spectrum of the LOS velocity averaged over the prominence is shown in the lower panel
of figure \ref{fig_power}.
We can recognize a distinct peak around 4 mHz ($P=$3 -- 6min) that corresponds to 
the dens stripes in figure \ref{fig_timeslice}.
Another peak is also seen around 1 mHz ($P=$15 -- 20min).
Again, these peaks are not recognized in the power spectrum of the H$\alpha$ intensity shown 
in the upper panel of figure \ref{fig_power}.
In the $k-f$ diagrams for the central horizontal slit shown in the left panel of figure \ref{fig_kfpower},
we recognize oblique ridges of enhanced power (indicated by arrows). 
The slope of the ridge $f/k$ suggests the phase speed of the 4 min wave about 100-200 km/s.

These properties are not specific for this particular prominence, but
we recognized that the periods of around 4 min and 15 min are quite common 
whenever we find the signatures of oscillations or propagating waves in prominences.
In all such cases, we do not find the corresponding periodic variation in the intensity.
This is exactly what was reported in Paper I for a prominence on 2022 September 26 
(panel \textit{f} of figure \ref{fig_images}),
although the period of waves have a certain distribution around 4 min and 15 min 
among the different prominences or even in one prominence.
It is also noted that preferable periods of 3 -- 5 min and 10 -- 20 min are
certainly consistent with the past studies by a number of authors as mentioned in section 1 and seen in table 1.
It is remarked here that, in a few cases, we observe a periodic variation in the central intensity
of H$\alpha$ line.
However, if we integrate the intensity over the line profile, such variation usually disappear.
Since, as is evident, the intensity of spectral line at a fixed wavelength point changes with the Doppler shift, 
the detection of intensity oscillation from observations at a single wavelength should be
regarded with a caution. 
We will refer the waves with a period of 3 -- 5 min and 10 -- 20 min 
as 4 min and 15 min waves, respectively, for the simplicity in this paper.
As suggested in Paper I, 
the 4 min waves are likely Alfv\'{e}nic waves propagating 
along the magnetic field supporting the prominence in the corona.

It is noted here that 
we recognized a clear signature or a hint of lower frequency oscillations with periods 
of 20 min -- 1 hour in about 30 prominences in our samples.
However the power spectrum density at low frequency is sometimes not reliable
due to the limited length of observed time sequences and it could be influenced by
time evolution of prominence structures or convective motions especially in hedgerow prominences.
By this reason, we focus on oscillations or waves with a period shorter than 20 min
in this paper.

\subsection{Grouping of prominences}

By scrutinizing the space-time diagrams for the central slits as shown in figure \ref{fig_timeslice}
and the averaged power spectral density as shown in figure \ref{fig_power} obtained for each prominence,
we classified the prominences in our dataset into four groups, 0, 1, 2 and 3,
by the presence of the short period waves with the following criteria;
\begin{enumerate}
\setcounter{enumi}{-1}
\item waves and oscillations are not recognized
\item there is a hint of oscillations  
\item oscillation or propagating wave are present with a confidence in a part of prominence and time duration
\item very clear signature of oscillation or propagating waves over the entire time period of observation
\end{enumerate}
Since it is difficult to obtain the confidence level of the power spectra through all prominences 
in automated manner and, 
in many cases, scrutinizing the space-time plot provides a more confident clue 
for capturing faint signatures of oscillations, 
the classification was made by human eyes of the four authors.
In this investigation, we recognized that the period of very clear oscillations or waves as classified in group 3 
are distributed in two particular ranges of period, i.e.,  3 -- 5 min and 10 -- 20 min.
Therefore, we defined the group 0, 1, 2 and 3 for each of 4 min and 15 min oscillations.
The prominence presented in figures \ref{fig_timeslice} -- \ref{fig_kfpower} is
an example that belongs to group 3 of both 4 min and 15 min oscillations.
It is remarked that, for prominences in the 4 min group 3,
we always recognized a clear signature of propagation of waves along the horizontal slit.
To give some ideas on how we classified the prominences into the four groups, 
examples of space-time plot of LOS velocity along the horizontal slit and averaged power spectrum
are presented for some prominences belonging to different groups in figure \ref{fig_appendix} of Appendix A2.
It is noted that we recognized oscillations with a period also in a range of 5--10 min 
in several prominences, but they are not persistent nor as conspicuous as those found in 
4 min and 15 min ranges in prominences of group 3.

For prominences in ether of group 3, we make further analysis to evaluate the phase speed 
and the direction of the wave propagation, which will be described in the following section.
For a few selected prominences in group 3, we also investigated EUV images at 193\AA\ or 171\AA\ 
taken by AIA/SDO in order to examine if there recognized corresponding waves in the corona.

\section{Results}\label{sec_result}

\subsection{Statistics on the presence of oscillations}\label{subsec_stat}

Table \ref{tab:4min_group} summarizes the number of prominences categorized 
to the groups 0,1,2 and 3 for the 4 min oscillation for each type of prominence.
Outstanding signatures of the propagating 4 min wave like those seen
in figures \ref{fig_timeslice} -- \ref{fig_kfpower} were found 
in 21 quiescent prominences out of 178 and they are categorized in group 3.
While at least a hint of the 4 min oscillation was found in 90 prominences (group 1$+$2).
In most cases the propagating features are found in space-time diagram along the horizontal slit.
Although the number of AR prominence is only three, 
it is notable that two of them are classified in group-2 of the 4 min oscillation.
Space-time diagram of the Doppler signal along the horizontal slit 
of one of the AR prominence is displayed in panel e of figure \ref{fig_appendix} in Appendix A2.

On the other hand, we found a very clear signature of the 15 min oscillation 
in 17 quiescent prominences and categorized them in group 3 of the 15 min oscillation, 
while at least a hint of 15 min oscillation was found in 60 prominences.
Table \ref{tab:corr_4_15} shows the number of prominences categorized to 4 groups 
of each 4 min and 15 min oscillations in a matrix form.
Although the number of prominences with a clear evidence or a hint of 15 min oscillation
is smaller than those of 4 min oscillation,
there is a certain correlation between the presence of 4 min and 15 min oscillations, i.e., 
prominences that have 4 min oscillation tend to have 15 min oscillation as well, and vise versa.
Therefore the waves in two ranges of periods are not likely independent phenomena
but have a common physical origin.
It is noted that the two AR prominences do not show any signature around 15 min period
and are categorized to group 0 of the 15 min oscillation.

\begin{table}
\caption{Number of prominences in 4 groups for the 4min wave}\label{tab:4min_group} 
\begin{center}
\begin{tabular}{cccccc}
\hline\noalign{\vskip3pt} 
 4-min group  & QR & PC & IM & AR & total \\   
\hline\noalign{\vskip3pt} 
3 & 17 (17) & 4 (4) & 0 (0) & 0 (0) &  21 (21) \\
2 & 33 (27) & 7 (2) & 1 (1) & 2 (2) &  43 (32) \\
1 & 34 (25) & 9 (5) & 3 (1) & 1 (0) &  47 (30) \\
0 & 36 (12) & 29 (11) & 2 (0) & 0 (0) &  67 (26) \\
\hline\noalign{\vskip3pt} 
total & 120 (81) & 49 (22) & 6 (2) & 3 (2) & 178 (107) \\
\hline\noalign{\vskip3pt} 
\end{tabular}
\end{center}
QR: quiescent prominence, PC: polar crown, IM: intermediate prominence, AR: active region prominence. 
Parentheses show the number of prominences with the horizontal size grater than 120arcsec.
\end{table}

\begin{table}
\caption{Correlation between the presence of 4-min and 15-min waves }\label{tab:corr_4_15} 
\begin{center}
\begin{tabular}{rrrrrc}
\hline\noalign{\vskip3pt} 
  15m group  & 0 & 1 & 2 & 3 & total \\   
\hline\noalign{\vskip3pt} 
4m group 3 &  2         & 4        & 6   	& \bf{9} & 21 \\
2 &  21       & \bf{10} & \bf{10}  & 2     & 43 \\
1 &  \bf{25} & \bf{12} & \bf{8}  & 2       & 47 \\
0 &  \bf{54} & 9        & 3        & 1       & 67 \\
\hline\noalign{\vskip3pt} 
total & 102 & 35 & 27 & 14 &  178 \\
\hline\noalign{\vskip3pt} 
\end{tabular}
\end{center}
Thick numbers are largest two values in each column. 
\end{table}

\subsubsection{Relation to the size of prominence}
\begin{figure}[t]
\centerline{
	\includegraphics[width=1.05\linewidth]{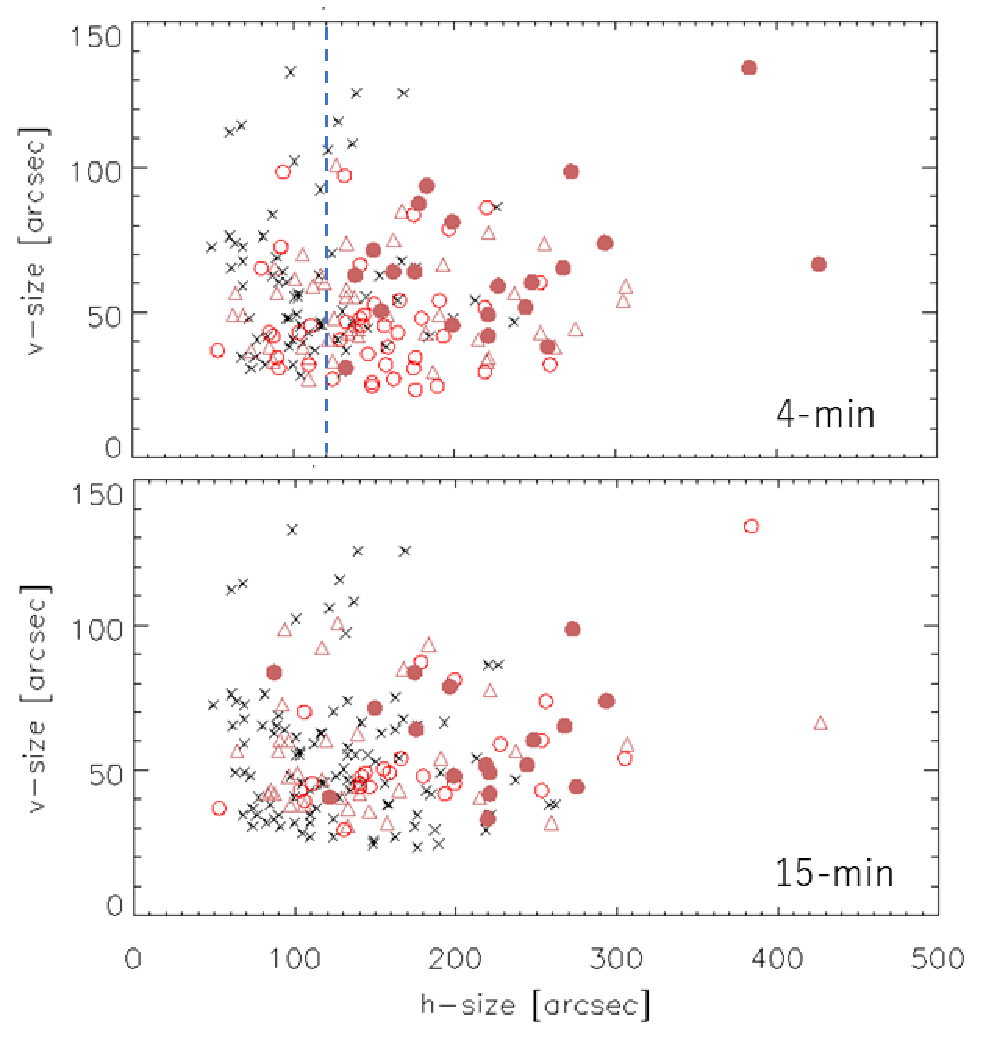}
}
\caption{Groups of prominences are shown with different symbols against their horizontal and vertical sizes. 
Upper and lower panels are for the groups of 4 min and 15 min oscillations, respectively. 
Black crosses, triangles, open circles and filled circles indicate prominences in group 0, 1, 2 and 3, respectively.
Vertical dashed line in the upper panel indicates the horizontal size = 120 arcsec.
{Alt txt: Two figures are arranged one above the other. 
X and y-axes of each panel are horizontal and vertical sized in arcsec and 
each prominence is marked by the assigned symbol to its group.
Upper panel is for the 4 min and lower panel is for the 15 min oscillation.  }
}
\label{fig_hvplot}
\end{figure}

Distributions of prominences in group 0$-$3 against their horizontal and vertical sizes
are shown in figure \ref{fig_hvplot} with different symbols for each group, 
i.e., crosses for group 0, triangle for group 1, open circles for group 2, and filled circles for group 3.
Upper and lower panels are for the group of 4 min and 15 min oscillations, respectively.
As for the 4 min oscillation (upper panel),
it is evident that clear signatures of the wave tends to be found in large prominences in their horizontal size.
The numbers in parentheses in Table \ref{tab:4min_group} give the number of prominence in each group 
of 4 min oscillation when we count only prominences with the horizontal size grater than 120 arcsec.
In this case, 21/107 $\sim$ 25\% of prominences show very clear evidence of the propagating 4 min wave
and 83/107 $\sim$ 78\% of prominences show at least a hint of 4 min wave or oscillation.
Therefore we may say at this point that the persistent wave in a period of 3 -- 5 min
propagating across the entire prominence
is not particularly rare but sizable fraction of large prominences harbor this phenomena.
The phase speed of the 4 min waves, if we clearly detected them in space-time diagram, 
is in a range of 100 -- 300 km/s.
Therefore the typical wavelength is about 200 km/s$\times$240 sec$\sim$50000 km or $\sim$70 arcsec.

The distribution of the 15 min groups shown in the lower panel of figure \ref{fig_hvplot}
tells us that the 15 min oscillation is also preferentially found in prominences with a large horizontal size,
but the tendency looks slightly less strong than the 4 min oscillation.
More significant difference from the 4 min oscillation is found in that the 15 min oscillation 
is rarely seen in low height ($<\sim$40 arcsec) prominences.
Also, in contrast to the 4 min oscillation, the propagating feature is usually not visible
in the space-time diagram for the horizontal slit.

The reason why small prominences tend to exhibit no clear signature of 4 min propagating waves
is partly due to the observational bias, i.e., the larger the prominence, the easier to detect the 
propagating wave because the size of small prominences is comparable or smaller than
the typical wavelength of 4 min oscillation.
But we think that this is not only the reason since, even in small prominences, 
oscillation without a clear evidence of propagation should be detectable.
Moreover prominences with a similar size show or not show the evidences of propagating waves
or oscillations.
This point is visualized in figure \ref{fig_images};  i.e., 
prominences $a$ -- $f$ are in group 3 for both 4 min and 15 min oscillations while $g$ -- $o$ are in group 0.
It is recognized that narrow and tall prominences are more common in group 0,
however some prominences in group 0 can have similar size and shape as those in group 3.

What is the condition by which a prominence harbors the short period (propagating) waves?
In the following, we investigate other aspects.

\begin{figure}[h]
\centerline{
	\includegraphics[width=1.1\linewidth]{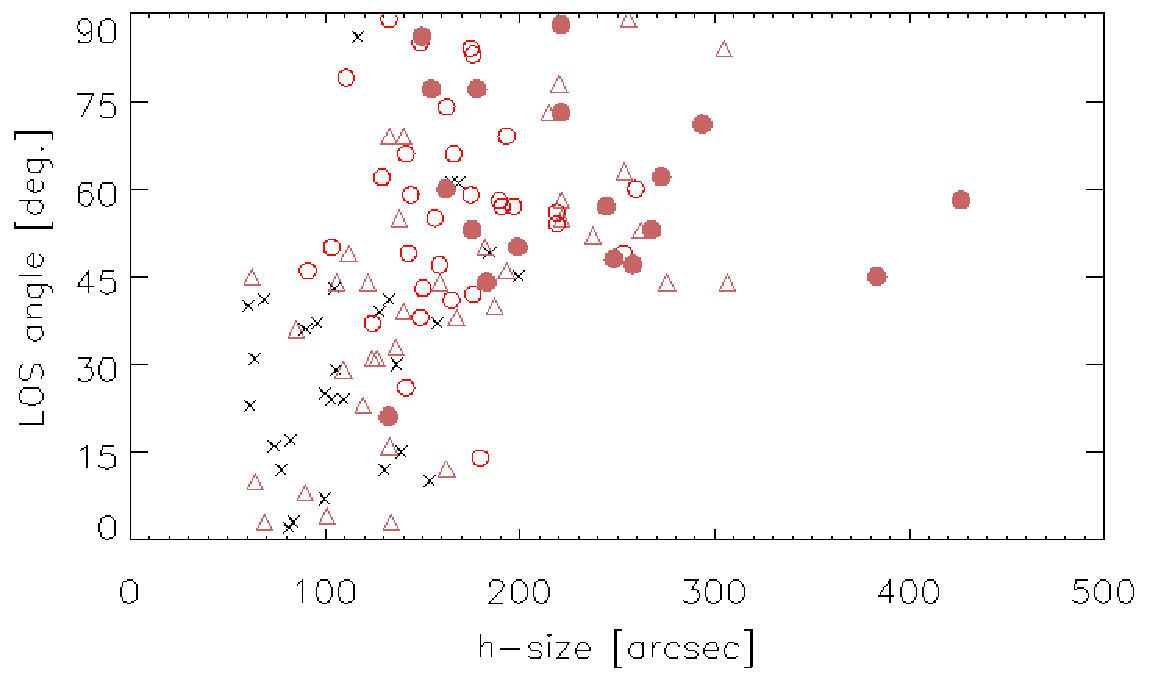}
}
\caption{LOS angle vs horizontal size of prominences with the same group symbols as in figure \ref{fig_hvplot}.
{Alt text: A figure displaying scattered symbols on a plane of horizontal size in arcsec and
LOS angle in degree.
Each prominence is marked by the assigned symbol according to its group.}
}
\label{fig_hlos_plot}
\end{figure}

\subsubsection{View angle of prominence}

One possible factor that affects the observed signatures of the propagating waves could be 
the angle between LOS and the axis of the prominence (i.e., dark filament on the disk).
Figure \ref{fig_hlos_plot} shows the distribution of prominences in 4 min group 0 -- 3
on a plane of LOS angle versus horizontal size with the same group symbols as before.
The LOS angle of 0$\degree$ and 90$\degree$ mean that we are looking the prominence 
from the direction of its axis and from a side, respectively.
The number of data points is reduced due to the limited measurement of the LOS angle
as explained in section 2.
A positive correlation is found between the LOS angle and the horizontal size of prominences,
This is naturally understood as the projection effect.
This correlation plot suggests that
the propagating 4 min waves is better visible when we observe the prominence from a side
and it implies that the 4 min waves preferentially propagate along the prominence axis.

Since it is also true that prominences with a similar size or LOS angle
show or not show the clear signature of the 4 min or 15 min waves,
we will further investigate other possible factors that relate the presence of the waves.

\begin{figure}[h]
\centerline{
	\includegraphics[width=1.1\linewidth]{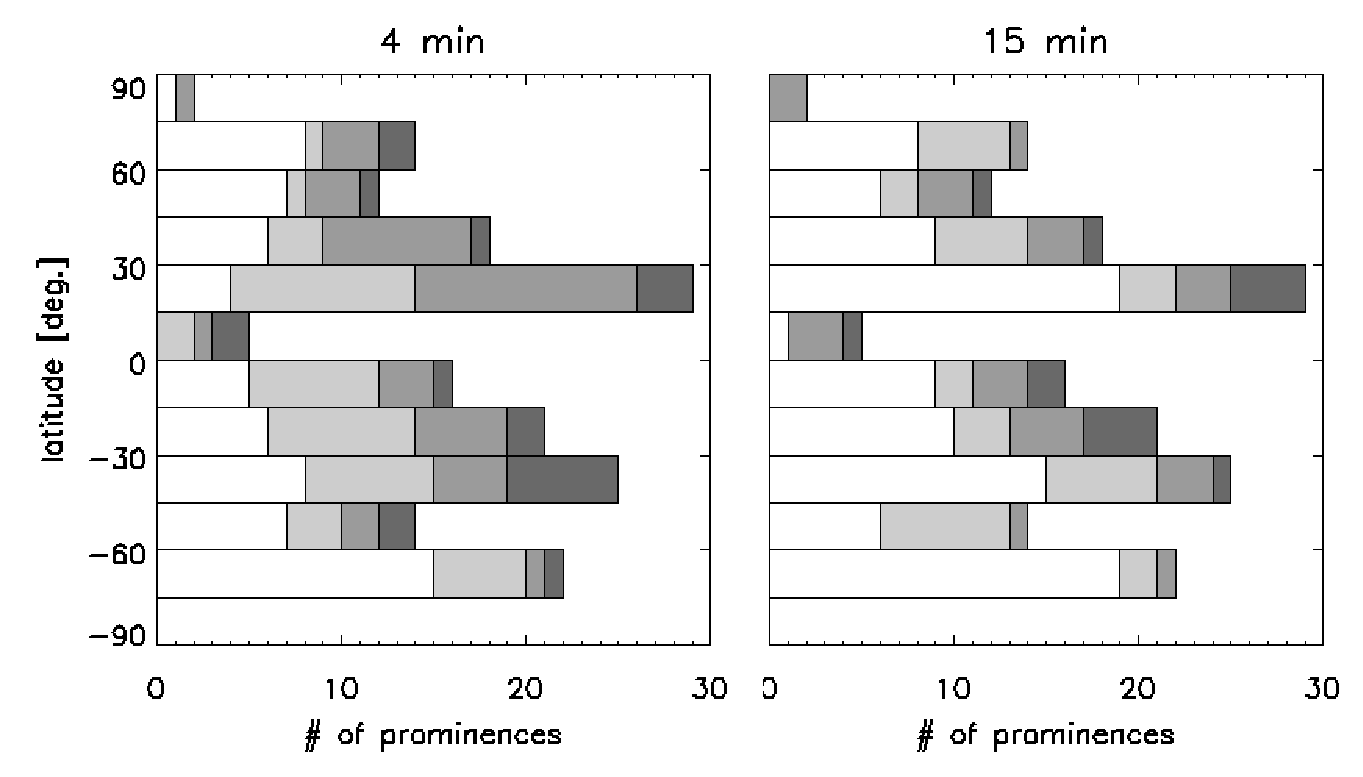}
}
\caption{Number of prominences against their latitude.
Left and right panels are for the groups of 4 min and 15 min, respectively.
In each bar, white, light gray, middle gray and dark gray parts represent 
the number of prominences in group 0, 1, 2 and 3, respectively.
{Alt txt: Two histograms are shown on the left and the right.
Each shows the number of prominences by horizontal length of bars against latitude
in vertical spanning from $-90$ degree to $+90$ degree with a step of 15 degree.
Each bar has white, light gray, middle gray and dark gray parts that represent 
the number of prominences in group 0, 1, 2 and 3, respectively.
Left and right panels are for the 4 min and 15 min oscillations, respectively.}
}
\label{fig_lathist}
\end{figure}

\begin{figure}[h]
\centerline{
	\includegraphics[width=1.05\linewidth]{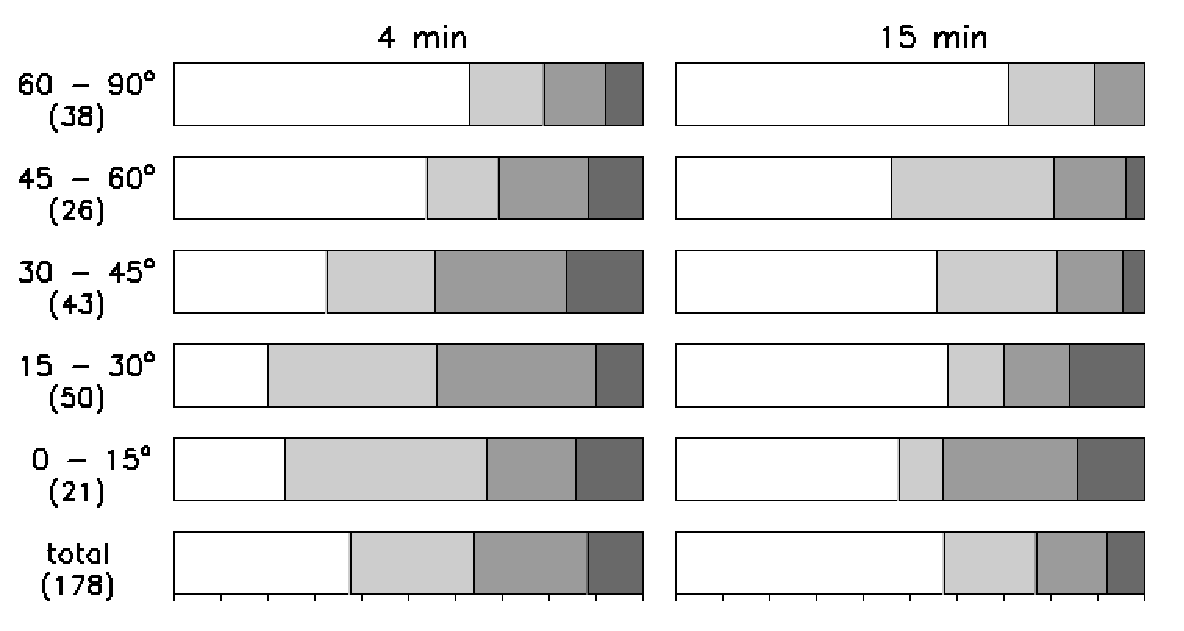}
}
\caption{Number ratio of prominences in each group for different latitude bands.
Bars in left and right are for the 4 min and 15 min groups, while
white, light gray, middle gray and dark gray parts in each bar represent the fraction of group 0, 1, 2 and 3, respectively.
The latitude intervals combining the northern and southern hemispheres are shown on the left, 
and the number of prominences in each latitude band are given by the numbers in parentheses..
{Alt txt: Two panels on the left and the right, each showing 6 horizontal bars  
arranged in vertical direction with a common length.
On the left, each bar is designated by the latitude intervals, i.e.,  from the to top the bottom,
as 60 -- 90, 45 -- 60, 30 -- 45, 15 -- 30, 0 -- 15 degree and total.
The fractional number of prominences in group 0, 1, 2 and 3 are indicated 
by white, light gray, middle gray and dark gray parts in each bar.
Left and right panels are for the 4 min and 15 min oscillations, respectively.}
}
\label{fig_latrate}
\end{figure}

\subsubsection{Relation to the magnetic activity}
As pointed in Paper I, the 4 min waves are likely Alfv\'{e}nic waves
propagating along the coronal magnetic field that supports the prominences, 
and therefore, the waves are possibly excited by external sources connected by the magnetic fields.
For example, outstanding oscillations observed in sunspots (umbral flush) could be
a candidate of the excitation source. 
In this regard, it is worth to see the environment where the prominences are located.
Figure \ref{fig_lathist} shows the histogram of the number of prominences against the latitude.
White, light gray, middle gray and dark gray parts in each bar represent 
the number of prominences in group 0, 1, 2 and 3, respectively,
and the left and right panels show those for the 4 min and 15 min oscillations.
These histograms tell us that a large number of prominences in our sample are
located in the low-middle latitudes in northern and southern hemisphere.
While in all latitudes, a certain number of prominences exhibit the 4 min and 15 min oscillations.
Figure \ref{fig_latrate} shows the fractional number of prominences that belong to the groups 0 -- 3
against the latitude bands combining the north and south hemispheres.
There is a hint that the fraction of prominences that show clear 4 min oscillation declines toward the high latitude.
However, this tendency could be a result from the fact that prominences in high latitude 
are often a polar crown filament extending in east-west direction, 
and they tend to be observed from the direction of their long axis on the limb.
In such case, the 4 min oscillation or waves are less visible as argued in the previous subsection.
Probably a more important message from these histograms is that a certain fraction of prominences 
harbor the 4 min and 15 min oscillations irrespective of their latitude.

To further confirm the relation between the prominence oscillations and surrounding magnetic activities,
we examined the distance of each prominence from the nearest active region and plage 
using the preceding or proceeding fulldisk H$\alpha$ images.
Although data is not shown here,
the result is negative, i.e., we did not fined an evidence of the relation between the 
presence of 4 min and 15 min oscillations in prominences with the distance to the nearest active region or plage.
From this result, 
the source of the short period waves should be attributed to ubiquitous disturbances 
present over the whole solar surface,
although the condition by which prominences harbor the short period waves or not is still unclear.

\subsection{Properties of the propagating waves}

In this subsection, we investigate the properties of the propagating waves
observed in prominences in group 3 in further detail.

\subsubsection{Periods}

\begin{figure*}[t]
\centerline{
	\includegraphics[width=0.9\linewidth]{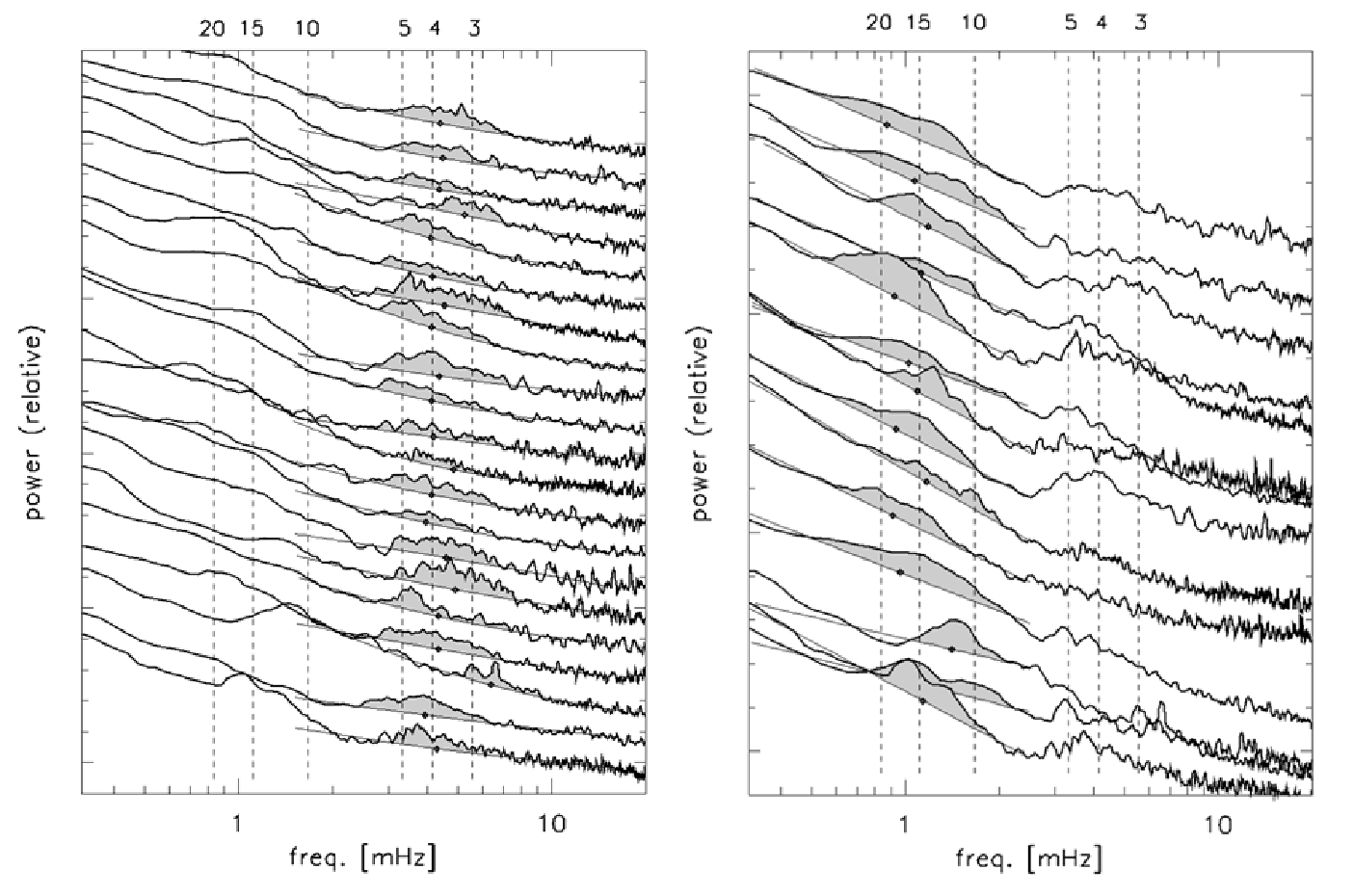}
}
\caption{Spatially averaged power spectrum densities of LOS velocity for prominences in group 3, i.e., 
21 profiles for 4 min oscillation (left) and 17 profiles for 15 min oscillation  (right).
Each power spectrum is displayed in logarithmic scale and successively shifted vertically to avoid overlapping.
Identified power enhancements and their central frequencies are indicated by gray areas and dots.
{Alt text: Two panels on the left and right both showing multiple power spectral density profiles
arranged in vertical direction.
X-axis is logarithm of frequency spanning from 3 to 20 mHz.
In the left panel, 21 profiles for group 3 of 4 min oscillation, and 
in the right panel, 17 profiles for group 3 of 15 min oscillation are plotted.
Enhancement of power around 4 min and 15 min periods are indicated by gray color in the left and right panels,
respectively.}
}
\label{fig_powg3}
\end{figure*}

Power spectrum densities of the LOS velocity spatially averaged over each prominence 
in group 3 are presented in figure \ref{fig_powg3}.
The left and the right panels are those for 21 prominences in group 3 of 4 min oscillation and
17 prominences in group 3 of 15 min oscillation.
Thin curve under each power spectrum is a 2nd or 1st order polynomial fit for the spectrum
in both sides of the 4 min or 15 min enhancement, respectively. 
Here the fitting is made in logarithmic scale of the period and the power,
and the intervals are 1.1 -- 11.1 min for 4 min and 6.7 -- 42min for 15 min,
while the region of enhanced power in the interval is excluded after some iterations.
Using this curve as the base of the power spectrum, we identify the enhancements 
of power around 4 min and 15 min and they are depicted by gray color in the left and right panels
of figure \ref{fig_powg3}, respectively.
Then we obtained the representative period of the 4 min and 15 min waves for each prominence
as the mean period weighted by the enhancement, which is marked by dots at the base of
each spectrum.
Also obtained are the approximate frequency ranges of the power enhancement.

\begin{figure}[t]
\centerline{
	\includegraphics[width=1.0\linewidth]{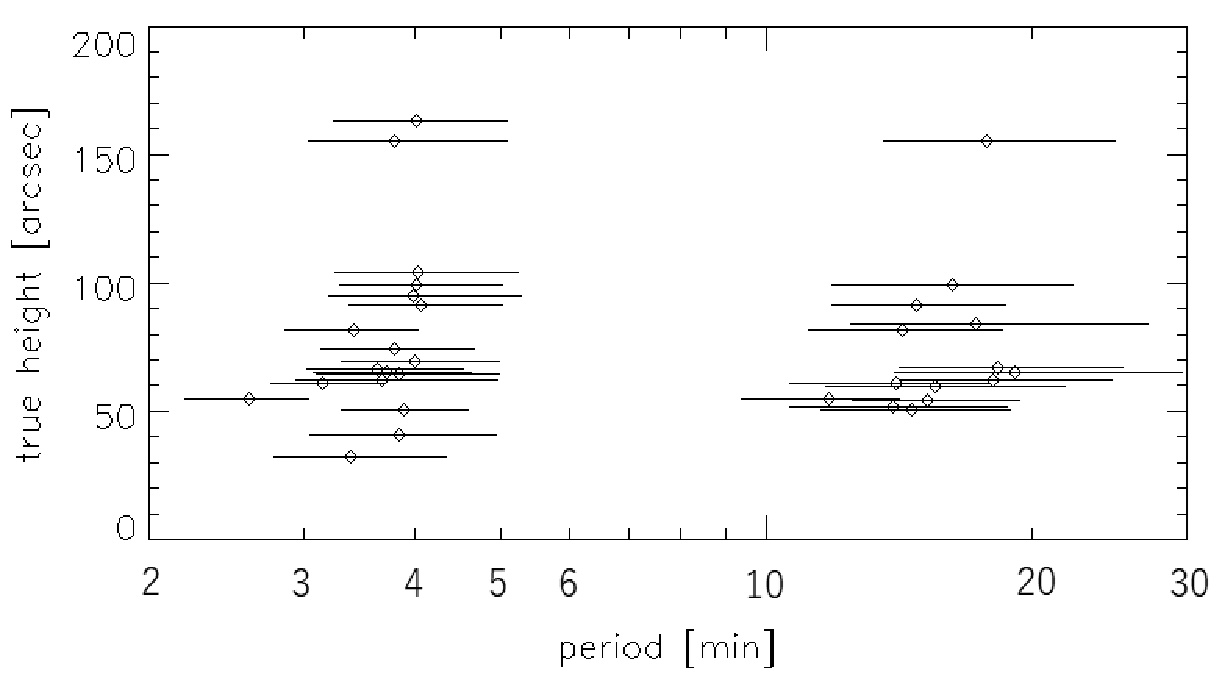}
}
\caption{Central periods of the oscillations determined for the prominences in group 3.
Vertical axis is the true height of prominences.
{Alt text: A graph displays multiple symbols with a horizontal line centered at each symbol.
The horizontal axis is logarithm of period ranging from 2 to 30 minutes 
and vertical axis is true height of prominence in arcsec.
Position of each symbol represents the central period of the oscillation and true height of prominences in group 3.
Horizontal lines indicate the width of power enhancement.
There are clusters of symbols in periods of 2 -- 5 min and 10 -- 20 min.}
}
\label{fig_period}
\end{figure}

Figure \ref{fig_period} displays the central period and range of power enhancement at the half peak
obtained for prominences in groups 3 of 4 min and 15 min oscillations together
by diamonds and horizontal lines centered at each symbol.
The vertical axis is the true height of prominences, which is obtained 
from the vertical size and the distance from the limb at the time of observation
using the equation (1) in \citet{Hashimoto_2026}.
In this plot, we first recognize that the central periods are clustered
in a range of 3 -- 5 min and 10 -- 20 min except one case, which has about 2.6 min.
Secondary, there are no significant correlation between the central period and the true height of prominences
in either 4 min and 15 min oscillations.
Also noticed is that the power enhancements do not have a sharp peak but extends in a certain range in frequency.
These results suggest that the oscillations are not a kind of single mode resonance in prominence bodies
because, if so, it would produce a sharp peak of power or a correlation between the period and their size.
The broadeness of the power enhancement could partly be due to the averaging of spatially variable period, 
but also due to superposition of traveling waves in slightly different periods.
Note that we do not find systematic variations of the central period of the power peak 
in its spatial distribution against the height or lateral position in a prominence.

\subsubsection{Phase velocity}

\begin{figure*}[t]
\centerline{
	\includegraphics[width=0.8\linewidth]{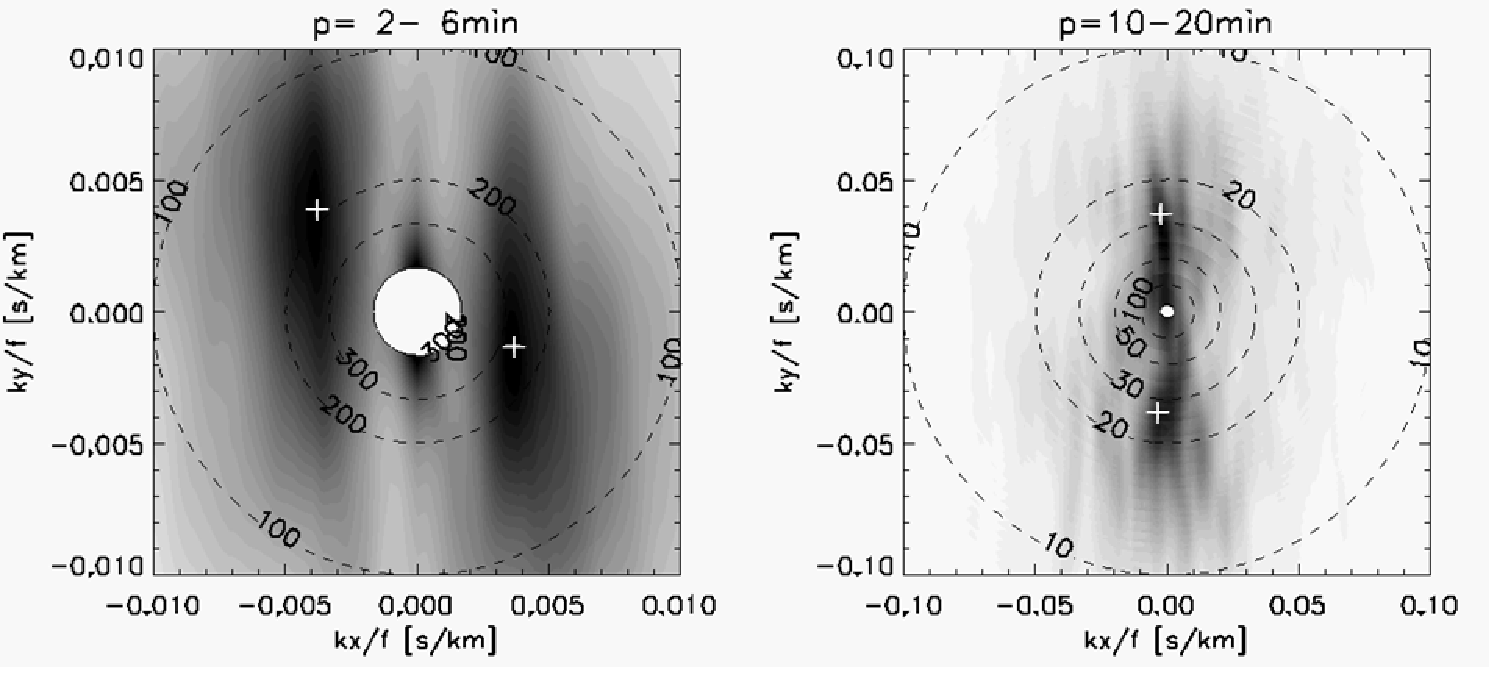}
}
\caption{Integrated power of the Doppler signal along lines of constant phase velocity in $(k_x,k_y,f)$ space.
Ranges of period for the integration are 2 -- 6 min in the left panel and 10 -- 20 min in the right penal.
Large power is displayed in dark.
{Alt text: Two gray scale maps on the left and the right.
Horizontal and vertical axes of each panel are kx/f and ky/f in sec/km and 
their ranges are from $-0.01$ to $+0.01$ in the left panel and from $-0.1$ to $+0.1$ in the right panel.
Dark regions represent a strong power integrated along lines of constant phase velocity in (kx, ky, f) space
in a range of period of 2 -- 6 min for the left panel and 10 -- 20 min for the right panel.
Peaks of power are marked by + signs.}
}
\label{fig_vppower}
\end{figure*}

\begin{figure*}[t]
\centerline{
	\includegraphics[width=0.8\linewidth]{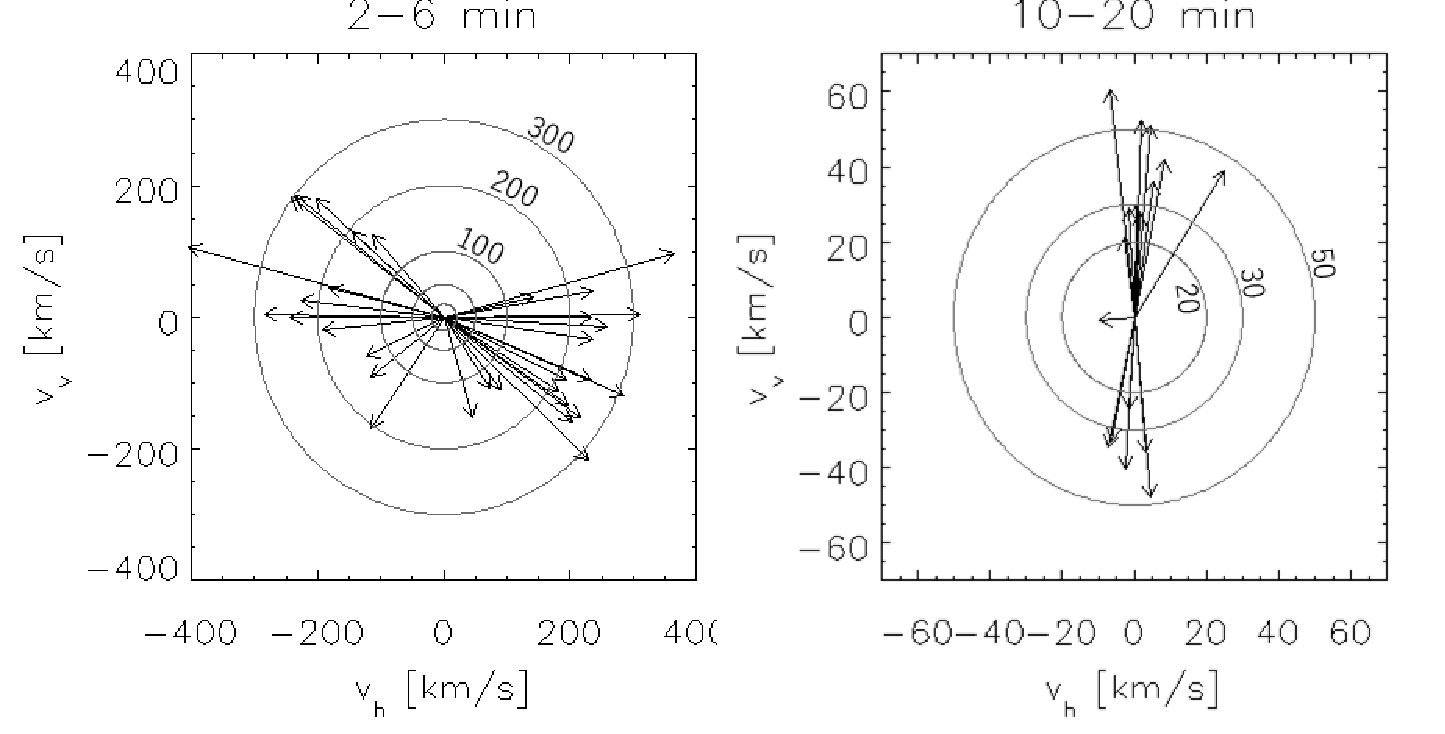}
}
\caption{Phase velocity of 4 min (left) and 15 min (right) waves in prominences in group 3.
Iso-speed lines are shown by circles accompanied by numbers showing the speed in km/s.
{Alt text: Two panels on the left and the right.
Horizontal and vertical axes of each panel are v-horizontal and v-vertical in km/sec.
Phase velocity of propagating waves are shown by arrows. 
Left panel is for the 4 min wave with a velocity range of $-400$ -- $+400$km/s.
Right panel is for the 15 min wave with a velocity range of $-70$ -- $+70$km/s.}
}
\label{fig_vparrows}
\end{figure*}

\begin{figure}[t]
\centerline{
	\includegraphics[width=1.0\linewidth]{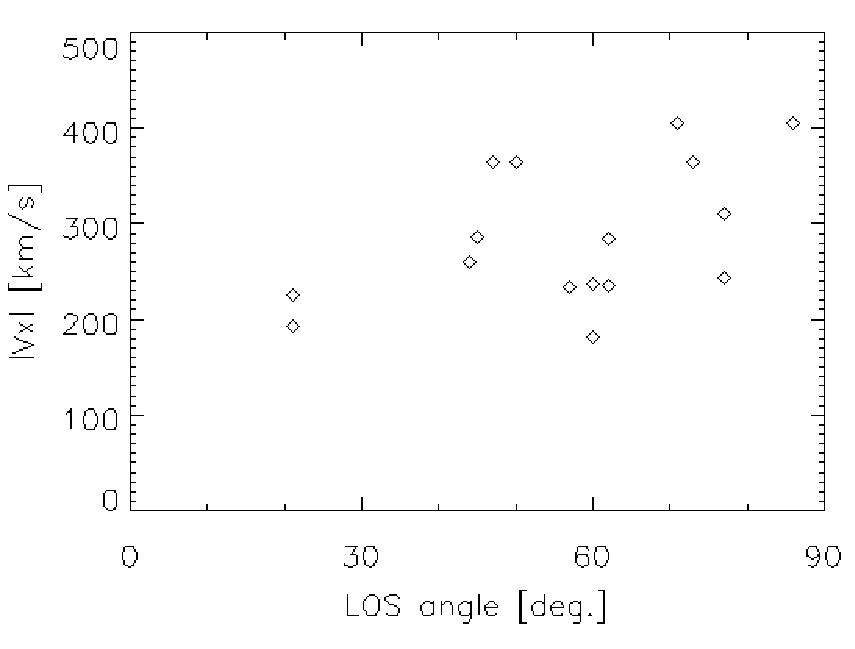}
}
\caption{Correlation between LOS angle and horizontal component of the phase velocity
of the 4 min wave.
{Alt text: A graph in which absolute values of the horizontal phase velocity of the 4 min wave
obtained for prominence of group 3 are plotted against the LOS angle by diamonds.
Horizontal axis, LOS angle, spans from 0 to 90 degree and the vertical axis,  absolute Vx, spans from 0 to 500 km/s.
Positive correlation between absolute Vx and LOS angle is seen.}
}
\label{fig_vpangle}
\end{figure}

The broadness of the 4 min power is also recognized in the power distribution on the $(k,f)$ plane 
shown in the left panel of figure \ref{fig_kfpower}
as an extension of the elongated power enhancement (ridge) 
along oblique lines of a constant phase velocity.
Therefore, the 4 min waves are not monotonic but a superposition of waves 
in different frequencies propagating with more or less a common phase speed.
The ridges of the power of the LOS velocity along a oblique line on the $(k,f)$ plane
are found also in other prominences in group 3 of 4 min oscillation
preferentially in the $k-f$ diagram for the horizontal slit.
Some prominences show more distinct and sharp ridges and some show less distinct ridges 
compared with those seen in figure \ref{fig_kfpower}.

It is important to notice that 
the slope of the ridge on the $(k,f)$ plane made for the horizontal slit is not necessarily 
the real phase speed of the wave in the plane of the sky
because the waves may not necessarily propagate along the horizontal slit.
If a plane parallel wave propagates in perpendicular to the slit, 
the power ridge will extends along the y-axis at $k=0$ and it indicates a infinite phase speed apparently. 
To exclude this uncertainty, we investigate the direction of propagation of the waves, 
by performing three dimensional Fourier transform of the LOS velocity in $(x,y,t)$ space
and obtained the power spectrum density in a three dimensional space $(k_x, k_y, f)$ for prominences in group 3.
Here $k_x$ and $k_y$ are wave numbers in horizontal and vertical directions.
If we see an enhancement of the power on the $(k_x,k_y)$ plane crossing at a frequency $f$ , 
it informs a presence of a wave propagating with a phase velocity $(f/k_x, f/k_y)$
projected on the plane of the sky.
Then, by integrating the power spectrum density along a line of constant phase velocity in $(k_x,k_y,f)$ space
in a specified frequency range,
we can obtain the power of the wave in that frequency range as a function of the phase velocity.
Before performing the Fourier transform, we extend the partial images of the LOS velocity 
to make their size common as 501$\times$501 pixels 
by adding null pixels surrounding the original Doppler images.
Since sampling of the spatial frequency in Fourier space is determined by the size of the image,
this treatment enables us to obtain the power spectrum density in $(k_x,k_y,f)$ space
with a common sampling in $k_x$ and $k_y$, and also among different prominences.
It helps us to obtain correct propagation vector from the $(k_x, k_y)$ diagram 
and to compare the power spectra of different prominences. 
It also improves the sampling of the spatial frequency although the true resolution 
is given by the original size of the images,

Figure \ref{fig_vppower} shows thus obtained integrated power on the $(k_x/f,k_y/f)$ plane
for the prominence on 2024 August 1 (the same prominence in figures \ref{fig_timeslice} -- \ref{fig_kfpower}).
The left and right panels are the results of integration for the periods of 2--6 min (4 min wave)
and 10 -- 20 min (15 min wave), respectively.
Contours of constant speed are shown by dashed circles.
Note that the scale of the axes is 10 times different between the left and the right panels,
i.e., we found that the phase speeds of the 4 min wave are about 10 times larger 
than those of the 15 min wave in all cases.
In the left panel of figure \ref{fig_vppower}, we recognize distinct power peaks in positive and negative $k_x/f$.
They represent the presence of waves with period around 4 min propagating rightward and leftward,
respectively, with a certain tilt inferred from $k_y/f$.
The peak positions of these enhancements, that are shown by plus marks, 
inform the representative phase velocity and its direction on the plane of the sky.
The power enhancements are elongated in vertical direction.
The reason of this
is due to the low resolution of vertical wave number because of the short height of prominence
or, possibly,  narrow spatial extension of wave channels in vertical direction.
The enhancements of  the integrated power 
are also seen in the result of the 15 min wave (right panel of Figure \ref{fig_vppower}).
In this case two peaks are found in positive and negative $k_y/f$ close to $k_x/f=0$,
which means that the 15 min waves are propagating in upward and downward directions in this prominence.

We measured the positions of the peaks of integrated power on the  $(k_x/f,k_y/f)$ plane
for prominences in group 3, and as a result,
we obtained the phase velocity vectors of 4 min and 15 min waves projected on the sky
for 21 and 12 prominences in group 3, respectively.
For some prominences we couldn't find isolated clear peaks, 
for some prominence we could identify only one peak, 
while for some prominences we identified more than two peaks 
in each integrated power map for 4 min and 15 min waves.
Left and right panels of figure \ref{fig_vparrows} show all phase velocities thus obtained  
with arrows on $(v_x, v_y)$ plane for 4 min and 15 min waves, respectively.
Here $v_x$ and $v_y$ are the reciprocal of $k_x/f$ and $k_y/f$, respectively.
It is clear that the 4 min wave is preferentially propagates in more or less horizontal direction with 
a phase velocity of 140--400 km/s, while the 15 min wave is in vertical direction with 
a phase velocity of 20--50 km/s.

Note that the vectors shown in figure \ref{fig_vparrows} are phase velocities projected 
on the sky plane and not true velocities in 3-D space.
Therefore if the phase velocity has a component in the LOS direction,
these results underestimate the true phase velocity. 
To test the projection effect, 
we examine the correlation between the phase velocity of the 4 min wave 
and the viewing angle of the prominence.
Figure \ref{fig_vpangle} shows the absolute values of the horizontal component of the phase velocities, $|v_x|$,  
against the angle of LOS to the prominence axis.
In this plot, only phase velocity vectors whose direction is close to the horizontal direction
within $\pm 30\degree$ are included.
We find a moderate positive correlation between the horizontal phase velocity and
the viewing angle of prominences, i.e., the horizontal phase velocity projected on the sky tends to
be larger when we observe the prominence from a right angle.
This supports the interpretation that the 4 min waves preferentially propagate along the prominence axis
as discussed in previous subsection. 
The scatter of the data points possibly implies the intrinsic difference of the phase velocity.

\section{Discussion}\label{sec_discussion}

\subsection{Nature of the waves in prominences}
In Paper I, we reported the presence of waves with periods 
around 4 min and 15 min in a prominence and that they 
propagate preferentially in horizontal and vertical directions across the prominence 
with a phase velocity of about 140 -- 200 and 24 km/s, respectively.
We interpreted the 4 min waves as the Alfv\'{e}nic wave propagating along the 
magnetic field supporting the prominence in the corona, while the 15 min wave as 
the Alfv\'{e}nic wave propagating in the prominence body.
The reason of this interpretation is the incompressible nature of the waves
suggested by the fact that there is no corresponding intensity variations
and the consistency of their phase velocity with the Alfv\'{e}n speeds 
estimated from the magnetic field strength obtained from 
the spectro-polarimetric data and the typical density of corona and prominences.
In the present work, although we don't have information of the magnetic field,
we found remarkably consistent characteristics of the waves in a number of quiescent prominences. 
It proves that such waves persistently propagating over the whole body of prominence
are not specific for the prominence studied in Paper I
but quite common phenomena.

Space-time diagram of the Doppler signal along the horizontal slit of one AR prominence
is shown in panel e of figure \ref{fig_appendix}.
As seen in the power spectrum density, this prominence has predominant period in 4 -- 6 min.
Remarkable is that the dens stripes in the space-time diagram showing this oscillation looks very steep,
namely near vertical.
Although we cannot determine the direction of propagating on the plane of sky due to the short height
of the prominence, if we assume that the wave is propagating in horizontal direction, 
the steepness of the stripes suggests a significantly high phase velocity of the propagation.
This is naturally interpreted by the Alfv\'{e}nic wave 
because active region prominences are supposed to have a stronger magnetic field.
It is also consistent with the result of \citet{Okamoto_2007}, i.e., they found
very high phase velocity of Alfv\'{e}nic waves in a horizontally stretched active region prominence
(see Table 1).

Waves with a period of 4 -- 5 min are so far observed also in Doppler signals of coronal emission lines 
by a number of observers 
\citep{Tsubaki_1977, Ichimoto_2000, Tomczyk_2007, Morton_2016, Morton_2025}.
Since these waves were not detected in intensity of the coronal emission lines, 
they are interpreted as incompressible, i.e.,  Alfv\'{e}nic waves propagating in the corona.
Indeed, as seen in the bottom panel of figure \ref{fig_timeslice}, we don't find a signature of 4 min wave 
in intensity of 193\AA\ taken by AIA in and around the prominence that showed conspicuous 4 min waves.
These coronal observations support the interpretation for the 4 min wave in prominences 
as the Alfv\'{e}nic wave propagating in the corona,
and, at the same time, 
it is suggested that prominences serve as a useful tracer of coronal Alfv\'{e}nic waves.

The nature of the 15 min oscillation seems to be less clear than the 4 min oscillation.
The positive correlation between the presences of 15 min oscillation and 4 min oscillation
as seen in Table \ref{tab:corr_4_15} 
may infer some hints about the origin of the 15 min wave, but the reason of which is not clear at this point.
As shown in table \ref{tab:corr_4_15}, 9 prominences in our sample show very clear signatures 
of both 4 min and 15 min waves propagating in different direction and phase speed.
If 15 min wave is an Alfv\'{e}nic wave propagating vertically in prominece body as suggested in Paper I, 
the 4 min and 15 min waves cannot take place in the same spatial volume
since magnetic fields cannot cross to each other.
Therefore these waves are necessarily propagating in different volumes superposed in the line of sight.
It is notable that \citet{Wiśniewska_2025} demonstrated 
the richness of wave periodicity and a superposition of long- and short-period kink-like oscillations 
in a quiescent prominence from a multi-wavelength analysis.

Another attractive interpretation of the 15 min waves is that they represent the 
cross-field slow propagation of MHD waves in a magnetic structure 
studied by \citet{Kaneko_2015}.
In this scenario, apparent propagation of oscillations is seen in LOS velocity when 
a magnetic arcade is observed from a side.
The phase speed is significantly slower than the Alfv\'{e}n speed in the corona, and
since the propagation is perpendicular to the magnetic field,
vertical propagation infer the predominantly horizontal magnetic field. 
Therefore the field configuration is consistent with that inferred from the 4 min wave,
and the positive correlation between the presences of 4 min and 15 min waves
could be understood as a geometrical effect.
Justification of this interpretation with a detailed comparison with MHD models is, 
however, beyond the scope of the present paper.

\subsection{Origin of the waves}
The nature of the 4 min wave as the Alfv\'{e}nic wave propagating in the corona
suggests that these waves are excited in external sources
and propagated along the magnetic fields that connect the prominence and the source regions.
The fact that the waves have rather broad spectra in frequency
may also suggest the origin of external sources, 
because, if the waves are of internal resonance trapped in prominences,
the power spectrum will have sharp or distinct peaks.
The broadness of the power enhancement is also the case for the 15 min waves,
and they are also likely propagated from external sources.

On the other hand, we didn't find evidence of connection between the presence of waves
in prominences and surrounding magnetic activities.
Therefore outstanding pulsations such as the umbral flush 
can be ruled out from the main candidate of the excitation source of the prominence oscillations.

\citet{Morton_2016} suggested that the short period (4--5min) Alfv\'{e}nic waves in the corona
could be excited by the mode conversion of the global p-mode oscillation at the transition region
or by the photospheric granular motions.
\citet{Wiśniewska_2025} suggested that the short period oscillations
in prominences originate from the 3 min oscillation in chromosphere.
Recently \citet{Chae_2024} found large amplitude transversal waves 
in chromospheric fine structures in a quiet region.
According to them, those waves have velocity amplitude up to 30 km/s and periods of 2--6 min.

Contrarily to the 4 min oscillation, 
periodic variations in the chromosphere in the period of 10 -- 20 min have rarely been reported so far.
Interestingly, \citet{Chae_2022} reported the presence of Alfv\'{e}nic waves in super-penumbra
around a small sunspot from a high resolution H$\alpha$ spectroscopic observation,
and they found that the waves are clearly separated into two groups of periods;
i.e., groups with a period in 1.9 -- 7.1 min and in 6.6 -- 22 min 
with the average of 3.7 and 12 min, respectively.
The coincidence of the two periods with those of our finding in prominences, 
i.e., 4 min and 15 min, is remarkable,
even though \citet{Chae_2022}'s finding is limited to a sunspot super-penumbra. 

In summary, 
our results of statistical study certainly suggest that the origin of the short period 
propagating waves in prominences is likely attributed to ubiquitous motions in solar atmosphere.
However definitive connections of these disturbances in chromosphere 
to the waves in prominences, and
the reason for the fact that some prominences do not show oscillation remain open questions.

\subsection{Diagnostics of the magnetic fields of prominences}
The clear signature of the propagating 4 min wave
tends to be found in quiescent prominences with a large horizontal size, i,.e., long filaments seen from a side.
This suggests that the magnetic field supporting the long quiescent filaments is predominantly
in horizontal and directed along the axis of prominences.
This picture is consistent with a number of past studies on the magnetic field of dark filaments
(e.g., \cite{Yamasaki_2023} and citations there in).
On the other hand, narrow and tall prominences seldom show such waves.
This implies that the propagation of the waves into a prominence
is influenced by the configuration of the magnetic field supporting the prominences.
The mechanism of such filtering is, however, remains as an issue for future works.

It is also true that a number of prominences don't show
any evidence of oscillation nor propagating waves in the Doppler signals
even though their size and shapes are not much different from the prominences that show very clear waves
(examples can be found in figure \ref{fig_images}).
A possible reason of this difference could be the polarization of the Alfv\'{e}nic waves,
i.e., if the wave oscillates linearly in the plane of the sky, they are not observed in the Doppler signal,
and, under our rather low spatial resolution, the transversal motions are not detected.

As mentioned in previous subsection, 
the co-existence of the 4 min wave propagating horizontally
and the 15 min wave propagating vertically may tells us a complex configuration 
of the magnetic field of prominences 
if both waves are interpreted as Alfv\'{e}nic waves propagating along the magnetic fields;
i.e., spatial volumes filled with magnetic fields in different directions are present along the line of sight
in a prominence observed on the solar limb.
On the other hand, if the 15 min wave is interpreted 
as the apparent cross-field propagation of Alfv\'{e}n waves in magnetic structure,
the combination of  observed properties of the 4 min and 15 min waves will provide
a powerful means for diagnosing the configuration of magnetic field supporting the prominence.

\section{Summary}\label{sec_summary}
Presence of small amplitude oscillations in solar prominences
were systematically investigated by analyzing 178 prominences on the solar limb.
It was found that more than half of prominences exhibit short period oscillations or waves 
in their Doppler signal,
and they are categorized into two distinct groups of periods,  
i.e., 3 -- 5 min (4 min) and 10 -- 20 min (15 min).
The 4 min wave is interpreted as Alfv\'{e}nic wave propagating in the corona 
preferentially in horizontal direction, 
while the nature of 15 min wave propagating preferentially in the vertical direction
is still controversial; two scenarios are discussed, i.e., 
Alfv\'{e}nic wave propagating in the prominence body 
and apparent cross-field propagation of Alfv\'{e}n waves in a magnetic structure \citep{Kaneko_2015}.
Since prominences that harbor the 4 min and 15 min oscillations are located in any latitude of the sun, 
the waves likely originate from ubiquitous motions in the chromosphere.
Recent high resolution observations begin to reveal the short period waves 
propagating in chromospheric fine structures (\cite{Chae_2024} and references therein).
While it is also true that a number of prominences do not show a signature of oscillations 
nor propagating waves, and the reason of this diversity remains open question.

Although we didn't derive the exact velocity amplitude of the short period waves 
in prominences in this study,
it is likely in a range of 0.5 -- 2 km/s as inferred from the paper I and previous works (see table 1).
If we assume the velocity amplitude of the waves as 1 km/s,  
spatial amplitude of 4 min and 15 min waves are about 100 km and 400 km, respectively.
These scales are well in the reach of current high resolution optical observations by large ground based telescopes.
Observations of LOS velocities by 
resolving lateral motions of individual threads in prominences, and, further more,
combining them with high sensitivity spectroscopic observations of surrounding corona 
to be performed by the forthcoming Solar-C/EUVST and MUSE 
will provide a vital clue for understanding the nature of these waves,
their excitation mechanism and energy transport in the corona.

\begin{ack}
The authors are grateful to Drs. S. Nagata, T. Iju, D.Cabezas and other members of Hida observatory 
for the daily operation and the effort on maintaining the SMART telescope. 
We also thank Dr. P. Antolin for valuable suggestions on the interpretation of our results 
with MHD models.
The SDDI was constructed under the support of MEXT/JSPS KAKENHI Grant Number JP15H05814.
This work is also supported by MEXT/JSPS KAKENHI Grant Number JP24K07093.
SDO is a mission of NASA’s Living With a Star Program.
\end{ack}

\appendix 
\section{Procedure for determining the view angle of the prominence axis}\label{app1}

In this appendix, we describe the procedure for determining the angle between LOS and 
the axis of the prominence.

\begin{figure}[h]
\centerline{
	\includegraphics[width=1.1\linewidth]{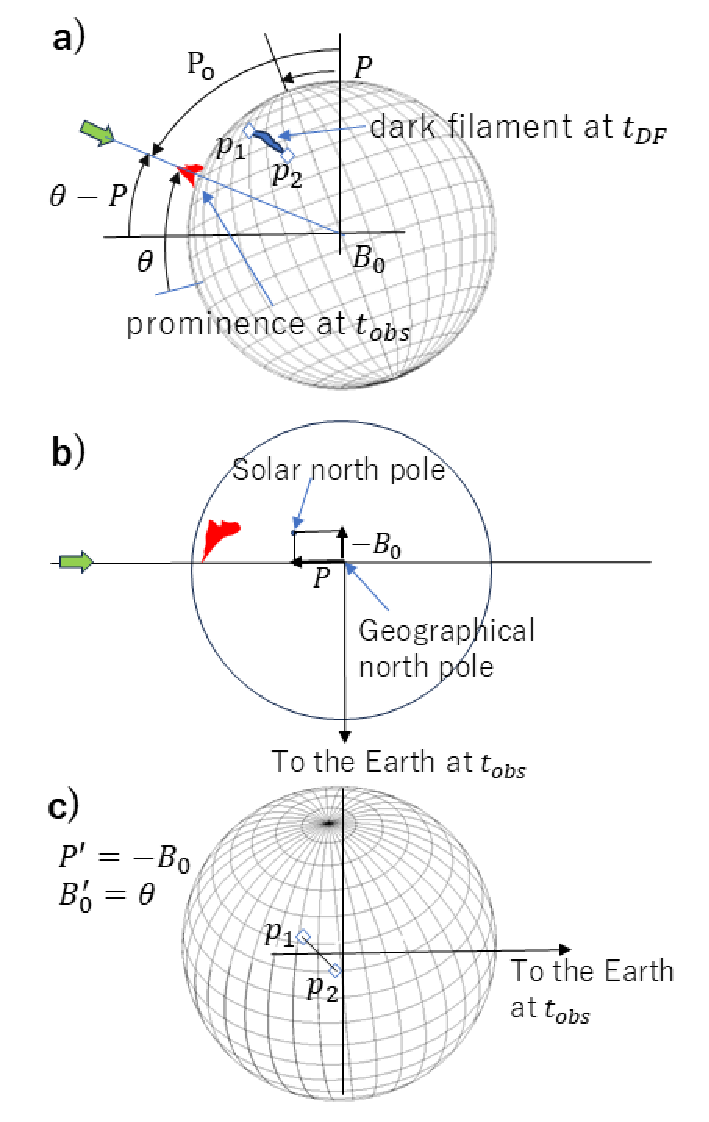}
}
\caption{From the top to the bottom, shown are the solar disk viewed 
from the Earth at the observations of a prominence (a), viewed from the geographical north (b)
and viewed from the top of the limb where the prominence was observed (c).
In panel (a), the prominence observed at $t_{obs}$ and the corresponding dark filament
observed at $t_{DF}$ are shown. 
{Alt text: Three panels each showing the solar disk seen from different directions are arranged vertically
and labeled as (a), (b) and (c) from the top to the bottom.
In panel (a), the target prominence is displayed on the upper left side of the limb, and corresponding
dark filament observed after a few days are shown in the upper left of the disk.
Panel (b) is the the disk seen from the geographical north and the position of the solar north pole is indicated.
Panel (c) is the disk seen from the top of limb where the prominence was observed.
Both ends of the filament corresponding to the prominence is marked by symbols p1 and p2 near the center.}
}
\label{fig_viewangl}
\end{figure} 

Panel (a) of figure\ref{fig_viewangl} shows the solar disk seen from the Earth 
at the time of the prominence observation ($t_{obs}$), 
where the position angle (angle between 
the solar north and the geographical north) and the central latitude of the Sun are $P$ is $B_0$, respectively.
The prominence on the limb at the polar angle (angle from the geographical north) $P_o$ and the 
corresponding dark filament observed in a following day at a time $t_{DF}$ are drawn.
On the H$\alpha$ image where the dark filament was observed, 
we select the position of both ends of the dark filament, i.e., $p_1$ and $p_2$.
We intend to make a map seen from the top of the limb where the prominence was observed
at the time of the observation, 
the viewing direction of which is shown by thick arrows on the left of panels (a) and (b).
Panel (b) of figure\ref{fig_viewangl} is the solar disk seen from the geographical north
with the direction to the Earth toward the bottom.
Therefore, the horizontal line crossing the center of the sun indicates the solar limb 
at the time of the prominence observation and 
the prominence is located behind the limb as seen from the Earth. 
In this map, the geographical north pole is at the center of the solar disk and
the solar north pole is shown by a dot together with the spans of $P$ and $B_0$.
It is recognized that a solar disk map seen from the top of the limb where the prominence was observed 
can be made with the position angle $P'=-B_0$
and $B_0'=\theta$, where $\theta$ is the latitude of the prominence given by $90\degree - P_o + P$.
The result of this transformation of the viewing direction is shown in panel (c) of figure\ref{fig_viewangl}, 
where plotted are the both ends of the dark filament, $p_1$ and $p_2$, 
at the time of the prominence observation which are obtained by rotating back the sun
using the time difference $t_{DF}-t_{obs}$ and the known differential rotation rate of the sun.
In this map, central meridian is the solar limb at the time of the prominence observation and
the horizontal line is the LOS of the observation.
Then the viewing angle of the prominence axis is determined from the angle between 
the line connecting $p_1$ and $p_2$ and the horizontal line.
At the same time we also obtained the distance of the prominence from the limb at the time of observation
as the distance of the center point of $p_1$ and $p_2$ to the meridian of the sun in degree.
 
\section{Examples of space-time plot and power spectra}\label{app2}

\begin{figure*}[h]
\centerline{
	\includegraphics[width=1.0\linewidth]{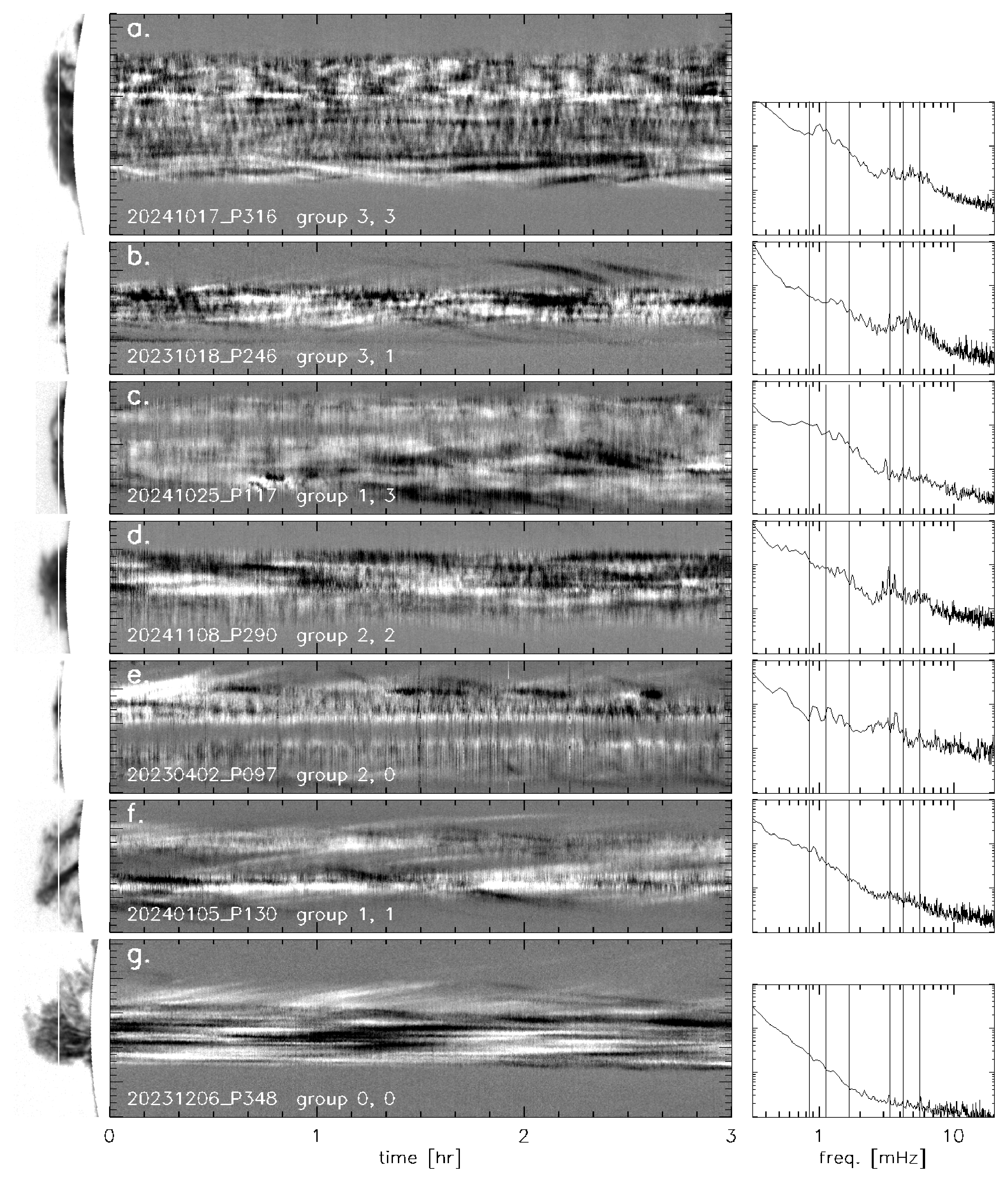}
}
\caption{Examples of space-time diagram and averaged power spectral density
for prominences belonging to groups 0 -- 3 of 4 min and 5 min oscillations,
From the left to the right, H$\alpha$ image of prominence with an artificial slit,
space-time plot of the Doppler signal for the horizontal central slit, 
and spatially averaged power spectral density.
At the bottom of each space-time diagram, group for 4 min and 15 min oscillations are shown.
Six vertical dotted lines in power spectrum on the right indicate, from the left to the right, 
the periods of 20, 15, 10, 5, 4 and 3 min, respectively. 
{Alt text: Seven rows are arranged in vertical.
Each row consists of, from the left to the right, H$\alpha$ image of the prominence with the limb at the right, 
space-time diagram of LOS velocity with horizontal axis be the time ranging 3 hours, and graph of
power spectral density against the frequency ranging from 3 to 20 mHz.
Each set of data is for a prominence belonging to the group of 4 min and 15 min of (3,3), (3,1), (1,3),
(2,2), (2,0), (1,1) and (0,0).}
}
\label{fig_appendix}
\end{figure*}

To give an idea on how we classified the prominences into the 4 groups of 4 min and 15 min oscillations,
some examples of the space-time plots and spatially averaged power spectra
are shown in figure \ref{fig_appendix}
for prominences belonging to groups other than the group 3.
Space-time plots along the central horizontal slit and power spectral density of only LOS velocity are shown
since they are the primary basis of our judgement.
Each space-time plot shows only for central 3 hours of time series
and shown on the left is the H$\alpha$ image at the beginning of this period.
The gray scale of the images is in negative.

\end{document}